%% file: paper.tex
\documentclass[
nofootinbib,
 amsmath,amssymb,
 aps,
 pre,
 longbibliography,
superscriptaddress
]{revtex4-2}

\usepackage[utf8]{inputenc}
\usepackage{graphicx}
\usepackage{dcolumn}
\usepackage{bm}
\usepackage{hyperref}

\usepackage{tikz}
\usepackage{tikz-cd} 
\usetikzlibrary{decorations.markings}
\usepackage{xcolor}
\usepackage{multirow}
\usepackage{tikz-cd} 

\newcommand{\ve}[1]{\bm{#1}}
\newcommand{\dif}{\mathop{}\!\mathrm{d}}

\newcommand{\eq}{\text{eq}}

\definecolor{mygreen}{rgb}{0.0,0.55,0.3}

\begin{document}

\title{Effects of phase separation on extinction times in population models}

\author{Janik Schüttler}
\affiliation{Department of Applied Mathematics and Theoretical Physics, University of Cambridge, Wilberforce Road, Cambridge CB3 0WA, United Kingdom}

\author{Robert L. Jack}
\affiliation{Department of Applied Mathematics and Theoretical Physics, University of Cambridge, Wilberforce Road, Cambridge CB3 0WA, United Kingdom}
\affiliation{Yusuf Hamied Department of Chemistry, University of Cambridge, Lensfield Road, Cambridge CB2 1EW, United Kingdom}

\author{Michael E. Cates}
\affiliation{Department of Applied Mathematics and Theoretical Physics, University of Cambridge, Wilberforce Road, Cambridge CB3 0WA, United Kingdom}

\date{\today}

\begin{abstract}
We study the effect of phase separating diffusive dynamics on the mean time to extinction in several reaction-diffusion models with slow reactions. 
We consider a continuum theory similar to model AB, and a simple model where individual particles on two sites undergo on-site reactions and hopping between the sites.
In the slow-reaction limit, we project the models' dynamics onto suitable one-dimensional reaction coordinates, which allows derivation of quasi-equilibrium effective free energies.  For weak noise, this enables characterisation of the mean time to extinction.
This time can be enhanced or suppressed by the addition of phase separation, compared with homogeneous reference cases.  We also discuss how Allee effects can be affected by phase separation.
\end{abstract}


\maketitle


\input{sections/intro}

\input{sections/chapter1}


\input{sections/chapter2}

\input{sections/conclusion}

\begin{acknowledgments}
We thank Tal Agranov for helpful discussions.
\end{acknowledgments}

\appendix

\input{sections/appendix}

\bibliography{refs_withdoi}

\end{document}

%% file: sections/intro.tex
\section{Introduction}

\subsection{Motivation}

A fascinating and important feature of stochastic systems is the ability to switch between different metastable states. For systems with an underlying equilibrium dynamics, which imparts a free energy structure, such transitions can usually be framed as the crossing of a barrier (typically a saddle) between two basins in the free energy landscape. Such crossings are rare events driven by noise (typically thermal) in the dynamics \cite{gardiner2009}. Similar events also occur in non-equilibrium systems, where they can be analysed by large-deviation theory~\cite{freidlin1998,touchette2009}, although the absence of a free energy makes the calculations more challenging in general.

Extinction events form an important class of such transitions, as studied in nonequilibrium statistical physics~\cite{lande1993,elgart2004, doering2005, assaf2010, meerson2011, doering2005, doubova2016, maciel2013}.  These are irreversible transitions in which a system changes from an `active' state, undergoing some nontrivial population dynamics, into a state of zero population. Since there is no recovery from such a state, these are examples of absorbing-state phase transitions \cite{hinrichsen2000}. 
Such transitions are relevant in a broad variety of fields including population biology \cite{lande1988,melbourne2008,ovaskainen2010,henle2004,davidson2009,purvis2000,saether2005}, epidemiology \cite{keeling1997,dykman2008,minayev2009,kamenev2008,allen2000} and climate modelling \cite{crowley1988,penn2022,feulner2009}. 
Understanding the factors that lead to extinction is of obvious importance, for example when trying to plan the global eradication of infectious diseases. Here typical questions include devising effective control strategies to facilitate extinction \cite{organization2009,keeling1997}.
Another highly pressing example is the loss of biodiversity caused by climate change \cite{cowie2022,urban2015,thomas2004,davidson2009,penn2022,feulner2009}. Here it is vital to understand the factors that influence the risk of extinction and possible strategies to optimally prevent species extinction \cite{ipcc2022}. In that setting, stochasticity in the dynamics can lead to extinction in any finite population, no matter how stable it appears to be. The likelihood of such an event is exponentially small for a large, apparently stable population but this also makes it exponentially sensitive to small changes in environmental control parameters, population fitness, or initial population size. 
 
The random noise that causes these rare events may arise internally (for example by demographic noise, such as random fluctuations in birth and death rates), or externally (for example environmental noise, leading to randomly time-dependent conditions)~\cite{lande1993,melbourne2008}.  
In this paper we focus on demographic noise, which is already sufficient to cause extinction, and we consider finite but large initial populations for which such noise is effectively weak and the mean time to extinction (MTE) becomes exponentially large \cite{assaf2010}.  Calculating the scaling of MTEs in such model systems is a long-standing challenge~\cite{lande1993,elgart2004, assaf2010, meerson2011, doering2005, doubova2016, maciel2013}; it can be achieved analytically for simple systems but it becomes more challenging as model complexity increases \cite{norris1997,doering2005}.
 
 An additional important aspect of these models is `the Allee effect' (defined more precisely for our purposes below).   This generally refers to the idea that fitness (or per-capita growth rate) is nontrivially dependent on population density, due to factors such as individual fitness, competition for resources, crowding effects, and transmission heterogeneity, which also influence the MTE \cite{lande1993,lloyd2007,bessa2004,drake2004,kramer2010}. 
 In some cases, the per-capita growth rate may have a maximum at some critical density below which cooperation makes fitness an increasing function whereas above it, competition makes it decrease \cite{stephens1999,dennis2002}. In its strongest form, the Allee effect can create a finite critical density below which extinction is almost certain. In this case, the MTE is given by the mean time for the density to fall to this critical value.
 
Given their status as rare events in stochastic many-body systems, MTEs are naturally studied using tools from statistical physics. However, these systems are generally far from equilibrium, so there is no single-valued, global free energy landscape.  Still, for systems with weak noise, one may define a quasipotential~\cite{freidlin1998,touchette2009} from which the asymptotic behaviour of barrier-crossing times can be determined.  The quasipotential is defined locally within each metastable basin and its computation is challenging in general.   However, 
if the dynamics of a non-equilibrium system can be reduced to a single Markovian reaction co-ordinate, the situation is simplified and it is often possible to reconstruct a global landscape representing an effective free energy.  We make extensive use of this fact in the following.

\subsection{Extinction coupled to phase separation and inhomogeneous mixing}
 
The main focus of this work is the effect of inhomogeneous densities on the MTE~ \cite{skellam1951}.  Extinction in systems that combine local population dynamics with spatial diffusion have been studied extensively, mostly focusing on the MTE and the most likely trajectory to extinction \cite{elgart2004, escudero2004, meerson2011, villamartin2015, doubova2016,agranov2021,maciel2013}.  However, there is so far little work on cases where spatial diffusion leads to spontaneous structuring of the population into regions of high and low densities. This is surprising since phase separation of this type represents a simple model of emergent heterogeneity at scales ranging from bacterial colonies \cite{cates2010,grafke2017} to human cities \cite{rogers2011}. Such spontaneous heterogeneity clearly has consequences for global population dynamics whenever the fitness is dependent on local population density. Less obviously, even without this coupling phase separation can alter the dynamics, because  the demographic noise terms are themselves density-dependent in general.

In the following, we characterise these effects in two types of model system, which are a field theory and a discrete particle model.  For simplicity, we work in a limit where the mixing (or phase separation) dynamics is fast compared to reactions (or birth/death).  This allows us to identify a one-dimensional reaction coordinate for the extinction transition, an approach that parallels previous works on nucleation in active matter \cite{cates2023}.  (We note that the extinction events considered here do not require that the population reaches exactly zero.  For example, the field theory considers a large volume $V$ and we treat sub-extensive populations as effectively extinct, under the assumption that recovery from such small populations is effectively impossible.  This point is discussed in more detail below.)

Within this setting, we extend previous results for MTEs, showing for the field theoretic-model that if phase separation arises for any global population density between its locally-stable state and zero, the MTE will be exponentially affected.  The discrete-particle model does not support phase separation because particles reside on just a few lattice sites, but we again find that inhomogeneous mixing of the particles has exponential effects on the MTE.
For both models, we also show how Allee effects can be modified by inhomogeneous densities.  In particular, we compare phase-separating field theories with suitable reference theories where the density remains homogeneous.  We find that the existence and type of Allee effect can be qualitatively altered by phase separation.

To preview the paper: Section~\ref{sec:modelab} analyses the field-theoretic models, which are of `Model AB-type'  \cite{li2020}.  They describe continuous density fields for a single species of particle, whose dynamics has a slow non-conserved part (similar to Model A~\cite{hohenberg1977}, describing birth-death processes or chemical reactions) and a fast conserved part (similar to Model B~\cite{hohenberg1977}, describing particle hopping, which drives phase separation).  The dynamics is far from equilibrium because, although the A and B dynamics each stem from an underlying free energy functional, these do not match. The Model A sectors of the chosen systems are inspired by well-known population dynamics models (the logistic and Schl\"ogl models \cite{murray2002,schlogl1972}), although we note that the noises in the field theories are Gaussian, as usual in models A and B~\cite{hohenberg1977}.  (The relevance and implications of non-Gaussian noises for rare events in these population dynamics models are discussed in Appendix~\ref{sec:discrete_ab}.)
The main results of Section~\ref{sec:modelab} concern the effect of phase separation on the MTE, and on the types of Allee behaviour that occur.

Section~\ref{sec:toy_model} describes the discrete-particle model, which is motivated as a scaled-down version of a full population dynamics model.  Two species of particles hop quickly between discrete sites and they undergo slow on-site chemical reactions.  Inhomogeneous mixing of the species is implemented by on-site interactions, where again the hopping and reaction dynamics are separately in detailed balance with respect to suitable free energies, but these do not match, driving the system away from equilibrium.  A weak-noise limit is obtained at low temperatures, so that extinction is a rare event. In addition to the effects of density inhomogeneity on the MTE, the simplicity of this model also allows computation of the paths to extinction, at the microscopic level, using transition path theory~\cite{metzner2009}.

Finally, Section~\ref{sec:conclusion} discusses our conclusions, including a summary of the main results, and a comparison between field-theoretic and discrete-particle models.  Some technical details are discussed in the Appendices.

%% file: sections/chapter1.tex
\section{Spatial Models for Population Dynamics with Phase Separation}
\label{sec:modelab}

This Section considers a class of stochastic continuum models in which the dynamics of a population is coupled to spatial phase separation. Previously, models of this general type have been used to describe, for example, the formation of bacterial colonies~\cite{cates2010,grafke2017}. A canonical example is ``Model AB'' as defined in \cite{li2020}, in which a scalar order parameter evolves via additive conserved (diffusive) and nonconserved (birth/death) processes. These are governed by incompatible potential functions so that, although each sector would separately represent a noisy gradient flow, together they do not. Equivalently, in thermal language, the conserved and nonconserved dynamics are governed by different free energy functionals such that, although detailed balance would be obeyed in each sector separately, it is broken in combination. 
The breakdown of detailed balance is a central feature in biological systems even at subcellular scales \cite{marchetti2013} and almost unavoidable at population-dynamics level \cite{okubo2001}.

In Model AB  \cite{li2020}, each free energy functional is chosen for simplicity to be of $\phi^4$, square-gradient form, with an order parameter field $\phi(\ve{x},t)\propto \rho(\ve{x},t)-\rho_0$, chosen so as to vanish at the critical density $\rho_0$ of the phase separation dynamics at weak noise. This is the density at which, under variation of other model parameters, the coexisting phases become identical. To study extinction, it is more natural to consider the $\rho$ field directly, in terms of which there is an absorbing state at $\rho(\ve{x}) = 0$ for all $\ve{x}$. We make this choice below, and also allow more general forms of birth-death dynamics than can be represented by a $\phi^4$ free energy. Nonetheless our spatial models remain of ``AB-type".
We emphasise that our theories employ Gaussian noises as usual in Models A and B~\cite{hohenberg1977}, although exact application of large deviation theories to chemical reaction or birth-death systems requires treatment of Poissonian noises that originate in the underlying discrete particles.  Implications of the Gaussian noises are discussed below in Sec.\ref{sec:discuss-AB}, and in Appendix~\ref{sec:discrete_ab}.

\subsection{AB-Type Models: General Framework}
\label{sec:ABtype}

We consider particles that are distributed in a large volume $V$ according to a coarse-grained density field $\rho(\ve{x},t)$ which evolves in time via both a non-conservative mechanism (Model A-type) and a conservative mechanism (Model B-type). Model A represents kinetic processes such as chemical reactions or population dynamics in which particle numbers change, while Model B represents diffusive relaxation of spatial structure, towards either a uniform or a phase-separated state. This terminology has long been standard in the statistical physics literature \cite{hohenberg1977}. Each dynamical sector has its own noise terms which are chosen following \cite{li2020}, such that detailed balance would hold in both the non-conserved and the conserved sector if taken in isolation. 

This allows chemical potentials $\mu_{A,B}(\rho)$ and local mobilities $M_{A,B}(\rho)$ to be identified for each sector, such that when the two are combined,
the density $\rho=\rho(\ve{x},t)$ evolves in time according to
\begin{align}
\label{eq:modelab}
\begin{split}
  \partial_t \rho &= -M_A \mu_A +  \sqrt{2\epsilon M_A} \Lambda_A  - \ve{\nabla} \cdot \ve{J}_B  \\
  \ve{J}_B &= -M_B \ve{\nabla} \mu_B + \sqrt{2 \epsilon M_B} \ve{\Lambda}_B 
\end{split}
\end{align}
Here $\Lambda_A$ is spatiotemporal unit Gaussian white noise with $\langle \Lambda_{A}(\ve{x},t) \rangle = 0$ and $\langle \Lambda_A(\ve{x},t) \Lambda_A(\ve{x}',t') \rangle = \delta(t-t')\delta(\ve{x} - \ve{x}')$, $\ve{\Lambda}_B$ is a vector of independent similar noises, and $\epsilon$ is the noise strength. Note that $\epsilon$ has been chosen equal for both sectors without loss of generality; this can always be achieved by appropriate rescalings of mobility and chemical potential. We are interested in the MTE for cases where extinction is a rare event, so we address the limit of small $\epsilon$ throughout. This has the advantage that the phase separation dynamics is accurately described by mean-field theory and it also means that It{\^o} and Stratonovich interpretations of the noises in \eqref{eq:modelab} are equivalent, despite the multiplicative noises.  We adopt It{\^o} calculus in the following, for concreteness.  Later it will emerge that, for the slow reaction limit addressed in this paper, the relevant noise scale for the extinction process itself is $\epsilon/V$. Hence it would be quite possible to have a rare extinction -- governed by a weak noise, large deviation limit -- in a system whose phase separation was strongly affected by noise, such that nontrivial exponents govern the behaviour of the B-type dynamics, close to its critical point~\cite{hohenberg1977}. However we do not address such cases here.

Following \cite{li2020}, we assume each chemical potential, $\mu_A$ and $\mu_B$, to derive from a separate free energy functional $\mathcal{F}_{A,B}$:
\begin{align}
\label{eq:ndjsakdnjqdqwqd}
  \mu_{A} &= \frac{\partial \mathcal{F}_{A}}{\partial \rho}, \qquad \mu_{B} = \frac{\partial \mathcal{F}_{B}}{\partial \rho}.
\end{align}
As such, the nonconserved or conserved dynamics, with the other switched off, would each lead to thermal equilibrium steady states with Boltzmann distributions
\begin{align}
  P_{A,\eq}[\rho] = \frac1{Z_A} e^{-\frac1{\epsilon}\mathcal{F}_{A}[\rho]}, \qquad   P_{B,\eq}[\rho] = \frac1{Z_B} e^{-\frac1{\epsilon}\mathcal{F}_{B}[\rho]} \,,
\end{align}
where $Z_{A,B}$ are partition functions. 
Importantly, the free energies $\mathcal{F}_A$ and $\mathcal{F}_B$ are not equal in general. The incompatibility of the free energy means that the dynamics \eqref{eq:modelab} is inherently out-of-equilibrium. 

Now assume that the A-type dynamics arises from an underlying birth-and-death process with birth rate $\lambda_b$ and death rate $\lambda_d$ per unit volume.  Reactions occur locally in space, so there is no particle transport caused by the A-type dynamics.  In this case the chemical potential and mobility are local functions of density, given in terms of the reaction rates by
\begin{align}
\label{eq:MA-muA}
\begin{split}
  M_A(\rho) &=  \frac12\left[ \lambda_b(\rho) + \lambda_d(\rho) \right],  \\
  \mu_A(\rho) &= 
   2\frac{\lambda_d(\rho) - \lambda_b(\rho)}{\lambda_b(\rho) + \lambda_d(\rho)}.
\end{split}
\end{align}
These two quantities determine both the scaling of the non-conservative noise in \eqref{eq:modelab} 
and the (average) reaction rate,
\begin{align}
\label{eq:RA}
 R_A(\rho) & = -M_A(\rho) \mu(\rho) \nonumber\\ & = \lambda_b(\rho) - \lambda_d(\rho) \; .
\end{align}
Note that all three functions $M_A,R_A,\mu_A$ are fully determined by the birth and death rates $\lambda_b,\lambda_d$.
We explain in Appendix~\ref{sec:discrete_ab} how these formulae can be motivated by applying a central limit theorem to the underlying birth-death process, in a suitable limit, which also determines the noise strength $\epsilon$.

Such a reaction system has a fixed point at any $\rho^*$ for which $R_A(\rho^*) = 0$. The fixed point is stable if $R_A' (\rho^*) < 0$ and unstable when $ R_A '(\rho^*) > 0$, where the primes denote differentiation.
In the absence of any conserved, B-type dynamics, there is no diffusion and hence no interactions between different locations in space. Thus the Model A sector equilibrates independently at every point in space. This implies that the free energy $\mathcal{F}_A$ can be written as $\mathcal{F}_A[\rho(\ve{x},t)] = \int\dif\ve{x}\, F_A(\rho(\ve{x},t))$ where the local free energy density $F_A$
depends on $\rho$ but not its gradients: 
\begin{align}
\label{eq:F_A}
  F_A(\rho) = \int\dif\rho\, \mu_A(\rho).  
\end{align}

Meanwhile, the conserved (Model B) part of the dynamics in \eqref{eq:modelab} is governed by the current ${\bf J}_B$ whose chemical potential $\mu_B = \delta \mathcal{F}_B/\delta \rho$ stems from a free energy chosen, following \cite{li2020}, as the standard Model B form
\begin{align}
  \mathcal{F}_B[\rho] &= \int\dif\ve{x}\,\left(F_B(\rho) + \frac12\kappa (\ve{\nabla}\rho)^2 \right)\,,  \label{eq:Bfree} 
  \\ 
  F_B(\rho) &= -\frac12\alpha (\rho-\rho_0)^2 + \frac14\beta(\rho-\rho_0)^4  \label{eq:Bfree-dens} 
\end{align}
with positive constants $\alpha,\beta,\kappa,\rho_0$.  The theory remains well defined for $\alpha<0$ but there is no phase separation in that case. As is usual in a Model B setting, the mobility $M_B$ is now chosen constant, unlike $M_A(\rho)$ for the reaction dynamics. There is no cubic term in $F_B(\rho)$ because the critical density $\rho_0$ is defined by the vanishing of $F_B''$ and $F_B'''$. Moreover, adding a linear term to $F_B$ has no effect on the dynamics because its contribution to $\mu_B$ is uniform in space.  

We define the global density as
\begin{align}
\label{eq:varphi}
  \varphi(t) = \frac1V \int\dif\ve{x}\, \rho(\ve{x},t)  
\end{align}
In the absence of the A-type birth-and-death dynamics, and assuming small positive noise parameter $\epsilon$, 
 the B-type dynamics in \eqref{eq:modelab} carries the system to a stationary equilibrium profile $\rho_{B,\eq}(\ve{x})$ which is the minimiser of $\mathcal{F}_B$ subject to a fixed total density $\varphi $:
\begin{align}
\label{equ:rhoB-min}
  \rho_{B,\eq}(\ve{x}) = \mathrm{argmin}_{\rho(\ve{x})} \mathcal{F}_B[\rho(\ve{x})]
\end{align}
It is important in the following that if the A-type dynamics is present but the timescales of A and B sectors are well enough separated that such reactions can be considered indefinitely slow, the B-type dynamics will always maintain the density profile at a solution of \eqref{equ:rhoB-min}, with the only time-dependence entering through the total density $\varphi(t)$.

\begin{figure}[t]
\includegraphics[width=\linewidth]{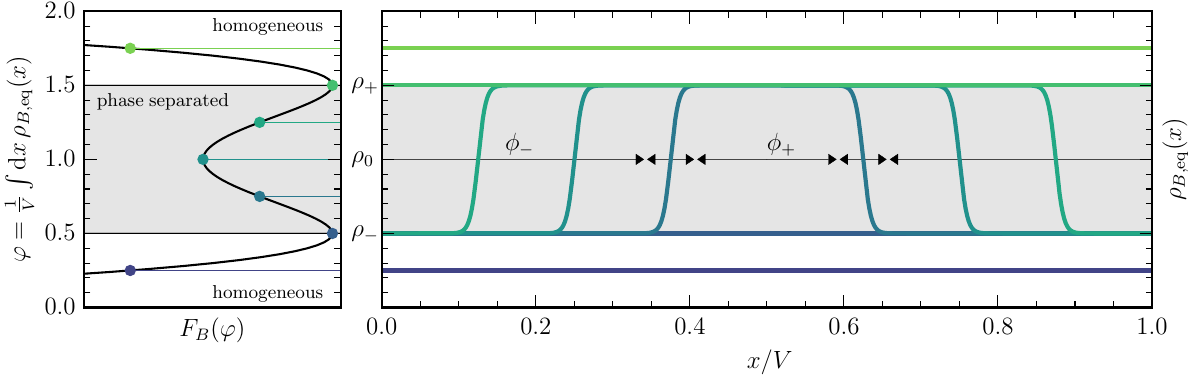}
\caption{Left panel: $F_B(\varphi)$ in Model-B (horizontal) as a function of the global mean density $\varphi$ (vertical). This is the free energy per unit volume $\mathcal{F}_B/V$ of a homogeneous state at density $\varphi$, which can be reduced by phase separation when $\varphi$ lies between the binodals, $\rho_ - < \varphi < \rho_+$. In the case shown, $\rho_\pm = 1 \pm 0.5$. Right panel: Resulting equilibrium profiles $\rho_{B,\eq}(x)$ in a one-dimensional box with periodic boundary conditions, for  $\varphi\in\{0.25,0.5,0.75,1.0,1.25,1.5,1.75\}$. For the phase-separated state with $\varphi = 0.75$, the phase volumes $V_\pm = \phi_\pm V$ of the dense and dilute phases are indicated via the black horizontal arrows, as are the sub-extensive interfacial `volumes'.}
\label{fig:modelb}
\end{figure}

Assuming that $V$ is large, we characterise the global minima in \eqref{equ:rhoB-min},
via a common tangent construction to find the convex hull of $F_B(\rho)$
as illustrated in Fig.~\ref{fig:modelb}.  The common tangent meets $F_B$ at the binodals densities $\rho_\pm = \rho_0 \pm \sqrt{\alpha/\beta}$ which in turn define the miscibility gap $[\rho_-,\rho_+]$. With the above choice of $F_B$, the tangent is $F_B' = 0$, this corresponds to a vertical line in the left panel of Fig.~\ref{fig:modelb}.
If $\varphi$
lies outside the miscibility gap, $\varphi \notin [\rho_-,\rho_+]$, the equilibrium density profile is homogeneous: $\rho_{B,\eq}(\ve{x}) = \varphi$ for all $\ve{x}$. In contrast,  whenever $\varphi \in [\rho_-,\rho_+]$ the equilibrium profile is spatially inhomogeneous, describing coexistence between two bulk phases at the binodal densities, whose phase volumes $V_\pm$ obey the `lever rule' whereby $V_-+V_+ = V$ and $V_-\rho_-+V_+\rho_+ = V\varphi$, encoding respectively the conservation of system volume and of matter. The relative volumes of the two phases $\phi_\pm:=V_\pm/V$ can then be written as
\begin{align}
\label{eq:volumefracs}
  \phi_+ = \frac{\varphi - \rho_-}{\rho_+ - \rho_-}, \qquad \phi_- = \frac{\rho_+ - \varphi}{\rho_+ - \rho_-}  .
\end{align}

When the system volume $V$ is not asymptotically large, the miscibility gap is narrowed slightly due to the sub-extensive square-gradient contribution in \eqref{eq:Bfree}. The resulting free energy cost takes the form of an interfacial tension $\sigma = (8\kappa\alpha^3/9\beta^2)^{1/2}$ at the interface between phases; this interface has width $\xi = (\kappa/2\alpha)^{1/2}$. Fig.~\ref{fig:modelb} illustrates such profiles schematically for various $\varphi$ values.  Allowing for the resulting `interfacial volume' one then has $\phi_++\phi_- = 1-\phi_\parallel$ with $\phi_\parallel\propto\xi V^{(d-1)/d}$. In what follows we consider $V$ large enough that \eqref{eq:volumefracs} can be applied uncorrected, effectively treating the interfaces as perfectly sharp.
Under periodic boundary conditions, the global mean density $\varphi$ and the binodal densities $\rho_\pm$ then entirely determine the equilibrium profile $\rho_{B,\eq}$, up to spatial translations, with no further dependence on the Model B parameters. 

Hence, assuming (as above) that the time scales for B-type dynamics are much faster than those of A-type reactions, the behaviour of the model in \eqref{eq:modelab} is fully determined by the density-dependent rates $\lambda_b,\lambda_d$, the (small) noise strength $\epsilon$, and the binodal densities $\rho_\pm$.

\subsection{Choices of Birth-Death Kinetics}
\label{sec:CBDK}

As representative examples for the birth-death rates, we consider two well-established models: the logistic model and the Schl\"ogl model \cite{murray2002,schlogl1972}. 
The logistic model can be interpreted as arising from the reaction system $P \rightleftharpoons 2P$, resulting in a volumetric birth rate $\lambda_b(\rho) = \rho$ and death rate $\lambda_d(\rho) = \rho^2 / \rho_s$. There is then a stable fixed point at the carrying capacity $\rho_s$, while the state of zero density $\rho = 0$ is an unstable fixed point.
If we denote stable fixed points by $\times$ and unstable fixed points by $\bullet$, the logistic model can be represented diagrammatically as
\begin{center}
\begin{tikzpicture}

\draw[line width=0.5pt] (0,0) -- (4, 0) ;
\node at (1.9, 0) {\Large $\times$};
\node[circle,fill=black,minimum size=5pt, inner sep=0pt] (c) at (0, 0){};
\node at (3.89, 0) {$\blacktriangleright$};
\node at (0.7, 0) {$>$};
\node at (1.3, 0) {$>$};
\node at (2.6, 0) {$<$};
\node at (3.2, 0) {$<$};

\end{tikzpicture}
\end{center}
The line represents density space $[0,\infty)$, and the arrows represent the gradient flow towards stable fixed points.

Since the death process $2 P \to P$ involves two particles but only removes one of them, the extinct state $\rho(\ve{x}) = 0$ for all $\ve{x}$ cannot be reached via $P \rightleftharpoons 2P$ in this logistic model.   To restore the possibility of extinction, we therefore assume an additional weak one-body decay process $P \to \emptyset$ which allows removal of the final particle and allows the population to reach extinction from states from states with $\varphi \simeq 1/V$.  Since we work at large $V$, this process is only relevant for very small $\varphi$, and does not enter our calculations below for the dynamics of the global density $\varphi(t)$, in which a single state $\varphi =0$ encompasses all configurations with subextensive particle number [$\varphi V = o(V)$].
 
Using (\ref{eq:MA-muA},\ref{eq:RA}), the logistic model has reaction rate and mobility
\begin{align}
\label{eq:logistic}
R_A(\rho) = \rho - \rho^2/\rho_s, 
\qquad 
2M_A(\rho) = \rho + \rho^2/\rho_s \; .
\end{align}
The dark blue line in the leftmost panel of Fig.~\ref{fig:allee} shows $R_A(\rho)$. The carrying capacity $\rho_s$ appears as a stable zero of $R_A$ with a second, unstable zero at the origin.  
Using \eqref{eq:F_A}, the free energy density $F_A$ for the logistic model is constructed by integration as 
\begin{align}
\label{eq:dsandqwhudehfgiobjfv}
  F_A(\rho) & = -2\int \frac{\rho\rho_s - \rho^2}{\rho\rho_s + \rho^2}\,\dif\rho
  \nonumber\\
  & = 2(\rho - \rho_s) - 4\rho_s \log\left(\frac{\rho+\rho_s}{2\rho_s}\right) 
\end{align}
which is shown as the dark blue line in the rightmost panel of Fig.~\ref{fig:allee}. 
The global minimum of $F_A$ occurs at the carrying capacity $\rho_s$ and the constant of integration was chosen in \eqref{eq:dsandqwhudehfgiobjfv} such that $F_A(\rho_s)=0$.

The Schl\"ogl model shares with the logistic model the feature of a carrying capacity $\rho_s>0$ such that death outweighs birth for $\rho>\rho_s$. However, the Schl\"ogl model requires a minimal population density for survival: there generally exists a positive critical density $\rho_u < \rho_s$ such that $R(\rho)<0$ for $\rho<\rho_u$, and the system is driven on average  towards extinction. The Schl\"ogl model therefore admits an unstable fixed point that lies between stable fixed points at the origin and at $\rho_s$. This situation is represented diagrammatically as
\begin{center}
\begin{tikzpicture}

\draw[line width=0.5pt] (0,0) -- (4, 0) ;
\node at (0, 0) {\Large $\times$};
\node[circle,fill=black,minimum size=5pt, inner sep=0pt] (c) at (1.3, 0){};
\node at (2.6, 0) {\Large $\times$};
\node at (3.89, 0) {$\blacktriangleright$};
\node at (0.7, 0) {$<$};
\node at (1.9, 0) {$>$};
\node at (3.2, 0) {$<$};

\end{tikzpicture}
\end{center}
The Schl\"ogl model has volumetric birth-and-death rates $\lambda_b(\rho) = (\rho_u + \rho_s) \rho^2$ and $\lambda_d(\rho) = \rho(\rho^2 + \rho_u\rho_s)$.  Hence
its reaction rate and mobility in (\ref{eq:MA-muA},\ref{eq:RA}) are
\begin{align} \label{eq:SMRM}
  R_A(\rho) = \rho(\rho - \rho_u)(\rho_s - \rho), 
  \qquad 
  2M_A(\rho) = \rho(\rho + \rho_u)(\rho + \rho_s)\,.
\end{align}
The right panel in Fig.~\ref{fig:allee} shows sample free energy functions for the Schl\"ogl model, given by \eqref{eq:F_A} as
\begin{align}
  F_A(\rho) &= 2\int\frac{\rho(\rho-\rho_u)(\rho-\rho_s)}{\rho(\rho+\rho_u)(\rho+\rho_s)}\,\dif\rho 
  \nonumber \\ & = 
  2(\rho - \rho_s) - 4 \frac{\rho_s + \rho_u}{\rho_s - \rho_u}\left(\rho_s\log\left(\frac{\rho+\rho_s}{2\rho_s}\right) - \rho_u \log\left(\frac{\rho+\rho_u}{\rho_s+\rho_u}\right)\right)\,.
\label{eq:Njndbashdwyd}
\end{align}
The unstable fixed point $\rho_u>0$ lies at the top of a free-energy barrier in $F_A(\rho)$. We again choose the integration constant via $F_A(\rho_s)= 0$. 

The Schl\"ogl model considered here can be realized microscopically by the reaction system $2P \rightleftharpoons 3P, P \to \emptyset$.
The original model of \cite{schlogl1972} included a reversible reaction $P \rightleftharpoons \emptyset$ which would lead to an additional constant term in $\lambda_b$, we exclude that possibility here, so that the model has an extinct state at $\rho=0$.

It is also useful to formally extend this model to allow negative values of the parameter $\rho_u$, specifically $0 < -\rho_u <\rho_s$.  We define $\rho_{us}=(-\rho_u\rho_s)^{1/2}$.  Then we extend the model by keeping $R_A$ as in \eqref{eq:SMRM} but modifying the mobility as $M_A(\rho) = \rho(\rho-\rho_u)(\rho_s-\rho)$ for $\rho<\rho_{us}$ while retaining $M_A$ as in \eqref{eq:SMRM} for $\rho>\rho_{us}$.
 In this case the diagrammatic fixed point structure reverts to that of the logistic model.  
 The resulting free energy is still given by  \eqref{eq:Njndbashdwyd} for $\rho>\rho_{us}$ while for $\rho < \rho_{us}$ we obtain from \eqref{eq:F_A} that
\begin{align}\label{eq:modifiedSM}
  F_A(\rho) = -2\rho + c
\end{align}
where the constant $c$ is chosen to make $F_A$ continuous at $\rho = (|\rho_u|\rho_s)^{1/2}$. 

This extended example underlines the fact that the free energy $F_A$ depends on both the mobility $M$ and the average rate $R$, which must respect physical constraints.  If, while taking $\rho_u<0$, one attempts to maintain the original noise variance $M_A$ from \eqref{eq:SMRM}, there would be range of density $\rho$ for which $M(\rho)<0$, which is clearly unphysical, from \eqref{eq:modelab}.

\subsection{Allee Effect and Definition of Allee Types}
\label{sec:Alleetype}

\begin{figure}[t]
\includegraphics[width=0.93\linewidth]{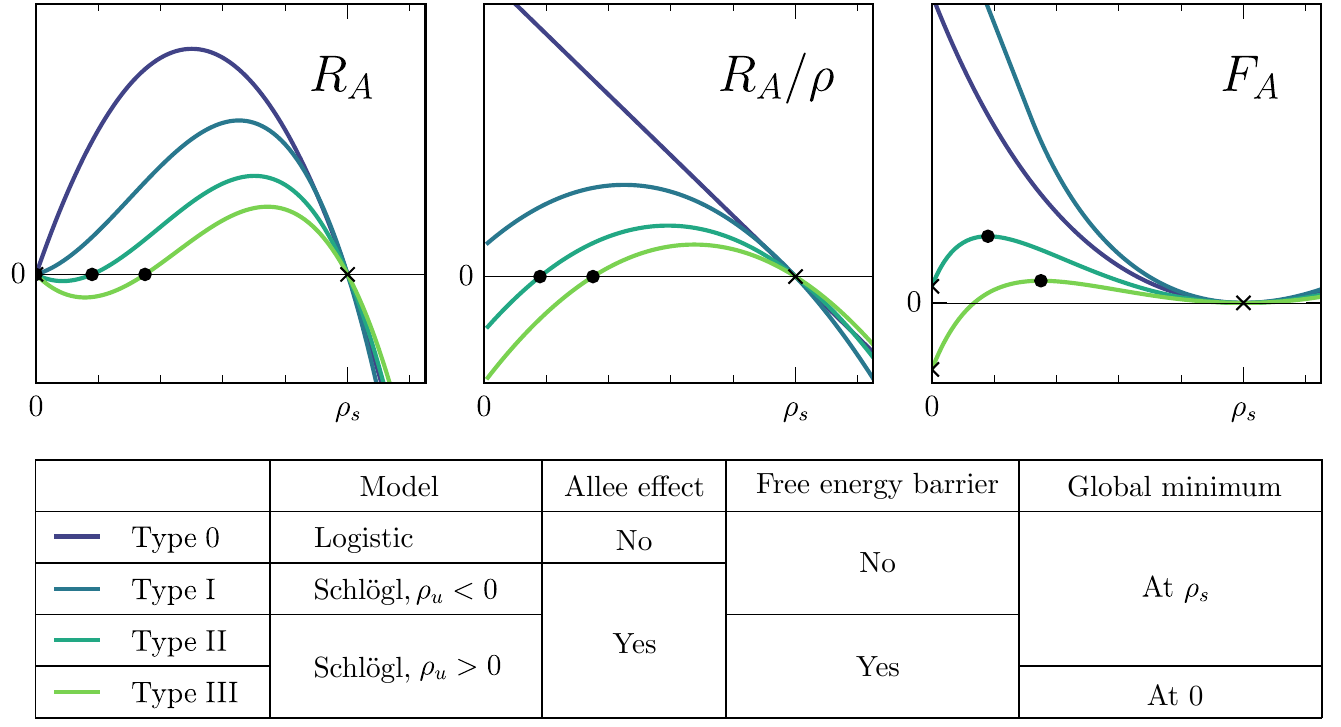}

\caption{Classification by `Allee Type' of the kinetic models considered in this paper. Rate function $R_A$ (left), per-capita rate function $R_A / \rho$ (centre) and Model A free energy $F_A$ (right). Stable fixed points are marked $\times$ and unstable fixed points $\bullet$. Type 0 is defined by the absence of an Allee effect, which implies a globally decreasing $R_A / \rho$. Types I--III have locally increasing $R_A / \rho$ at $\rho=0$ and thus exhibit an Allee effect. In Type I there is no unstable fixed point. Types II and III have such a fixed point and are distinguished from each other by whether $F_A(0)$ lies above or below $F_A(\rho_s) = 0$. 
}
\label{fig:allee}
\end{figure}

In population dynamics, an interpretation of the Schl\"ogl model involves cooperation among individuals, so that their fitness  increases with increasing population density at low enough population densities \cite{gastner2011}.  Here, the fitness is identified with the mean rate of population growth per capita, that is $R(\rho)/\rho$.
Examples of co-operative effects are access to potential mates, group hunting, and defending from predators. The model parameter $\rho_u$ then represents a critical density below which the population ceases to be sustainable and tends to shrink deterministically. 
Such cooperation is one example of an Allee effect \cite{stephens1999,dennis2002,gastner2011}. 
More generally, the Allee effect is a loosely defined concept in population dynamics describing the existence of a maximum in the individual fitness as a function of density $\rho$. Such a maximum arises because positive cooperation (or a similar mechanism) dominates at low densities while negative competition (crowding, etc.) dominates at high ones, as it does in the logistic model. 

In the models considered here, we use $R_A(\rho)/\rho$ is a proxy for individual fitness, and we say that an Allee effect occurs whenever this is nonmonotonic in $\rho$.  The centre panel in Fig.~\ref{fig:allee} shows various $R_A(\rho)/\rho$ curves for the logistic and Schl\"ogl models. In the logistic case there no Allee effect: here crowding decreases fitness even at the lowest densities, and indeed $R_A/\rho$ decreases linearly from the origin towards $\rho_s$. On the other hand, the curves shown for various parameters in the Schl\"ogl model each have a maximum in the per-capita reproduction rate and therefore can be said to have an Allee effect.
 
In our later discussions, it will be useful to divide the Allee effect into various types, starting with Type 0 where there is no Allee effect ({\em e.g.}, the logistic model). Type I has a maximum in $R_A(\rho)/\rho$, and hence an Allee effect as usually defined, but only one zero in $R_A(\rho)$. The fixed point structure is then indistinguishable from Type 0; in both cases there is no barrier in $F_A(\rho)$. This means that the path to a rare extinction event is `uphill all the way' on the free energy landscape.  Models of Type I can be found in the literature, for example a modified logistic model that incorporates an Allee effect \cite{dennis1989}. The extended Schl\"ogl model introduced above [with $\rho_u<0]$ is also Type I and is the example shown in Fig.~\ref{fig:allee}. 

Allee types II and III both have a free energy maximum ({\em i.e.}, a barrier) at density $\rho_u$; they are distinguished by the sign of the difference in free energy between the stable fixed points at $\rho = 0$ and $\rho=\rho_s$. Specifically, we define Type II as having $F_A(0)>F_A(\rho_s)$ and Type III as having $F_A(0) < F_A(\rho_s)$. The transition from II to III thus involves a jump of the global minimizer of $\mathcal{F}_A$ from a state of uniform density $\rho_s$ to the state of zero density. If the zero-density state was not in fact an absorbing one, the system would reach a stationary state. For Type II the stationary probability density would then be concentrated on global densities $\varphi \simeq \rho_s$, with rare excursions to $\varphi\simeq 0$, whereas the opposite behaviour would arise for Type III. Note that, in contrast to our other definitions of Allee types, the distinction between II and III depends on the noise structure as well as the mean reaction rate $R_A(\rho)$. 

In the literature, Type I is often referred to as a weak Allee effect, and Types II/III as a strong Allee effect \cite{stephens1999,dennis2002}. Other works have referred instead to Allee Types I/II as a weak Allee effect, and used a definition similar to Type III for the strong Allee effect \cite{meerson2011}. In the context of rare extinction events, the most important distinction lies between 0/I and II/III: for Types 0/I the extinction pathway consists of an excursion of ever decreasing probability to smaller and smaller densities, eventually reaching zero, whereas for Types II/III, the pathway comprises an improbable excursion to reach the unstable fixed point at $\varphi=\rho_u$, beyond which the population is driven to extinction deterministically. In the latter case, the extinction rate is set by the first passage time to reach $\rho_u$.

\subsection{Timescale separation and the Langevin equation for the total density}
\label{sec:projection1}

In this Section, we study the model defined in Eq. \eqref{eq:modelab} in the limit of slow reactions. 
As already noted in Sec.~\ref{sec:ABtype}, this time scale separation ensures that the density field
is always close to the minimum of the B-type free energy 
\begin{align}
\label{eq:approx}
  \rho(\ve{x}) \approx \rho_{B,\eq}(\ve{x})\,,
\end{align}
with $\rho_{B,\eq}$ given by \eqref{equ:rhoB-min}.
For self-consistency, 
we assume small but nonzero $\epsilon$: small enough that fluctuations of $\rho$ away from $ \rho_{B,\eq}$  are negligible, but nonzero so that, on the very slow timescale of reactions, any nucleation processes required to achieve phase separation have occurred. (Such nucleation is required in parts of the miscibility gap where $F_B''(\rho)>0$ so that the uniform state is metastable.)

We also recall from Sec.~\ref{sec:ABtype} that for large volumes we can consider the density $\rho_{B,\eq}$
to comprise two bulk phases of known compositions $\rho_\pm$ and phase volumes $\phi_\pm$, with negligible interfacial regions.  We show in this section that this results in a Markovian dynamics for the global density $\varphi(t)$, in which the overall birth and death rates are obtained by summing separately the contributions from the two phases.  While the Model AB system does not have detailed balance microscopically, the single reaction co-ordinate $\varphi$ becomes an autonomous degree of freedom that behaves as a  noisy gradient flow in a one-dimensional effective free energy landscape that we denote in the following by $F(\varphi)$.  Note that this $F$ is closely related to the quasipotential for the nonequilibrium problem, see Section~\ref{sec:discuss-AB}.

To derive a dynamical equation for $\varphi(t)$ and from it the effective free energy $F(\varphi)$, 
we begin by integrating both sides of the equation of motion \eqref{eq:modelab} over spatial coordinates. Dividing by the volume $V$ we have
\begin{align}
\label{eq:ndsajkdnqjwdq}
  \partial_t \varphi &= \frac1V\int\dif\ve{x}\, R_A(\rho(\ve{x},t)) + \frac1V\int\dif\ve{x}\, \sqrt{2\epsilon M_A(\rho(\ve{x},t))} \Lambda_A(\ve{x},t) - \frac1V\int\dif\ve{x}\, \ve{\nabla} \cdot \ve{J}_B\,.
\end{align}
The last term on the right-hand side vanishes because the Model B dynamics conserves the density. 
We now use \eqref{eq:approx} to approximate the remaining two terms
\begin{align}
\label{eq:ndsajkdnqjwdq2}
  \partial_t \varphi &\approx \frac1V\int\dif\ve{x}\, R_A(\rho_{B,\eq}(\ve{x},t)) + \frac1V\int\dif\ve{x}\, \sqrt{2\epsilon M_A(\rho_{B,\eq}(\ve{x},t))} \Lambda_A(\ve{x},t)\,.
\end{align}
The next steps depend on whether or not $\varphi(t)$ lies in the miscibility gap. If not, the system is homogeneous so $\rho_{B,\eq}(\ve{x},t) = \varphi(t)$ and the first term on the right hand side becomes 
$R_A(\varphi(t))$. Likewise, by summing independent Gaussian variables, the second term is
\begin{equation}
\frac1V \int\sqrt{2\epsilon M_A(\varphi(t))} \Lambda_A(\ve{x},t)\,\dif\ve{x} 
= 
\sqrt{\frac{2\epsilon}{V} M_A(\varphi(t))}\, \Lambda(t)
\end{equation}
where $\Lambda(t)$ is a purely temporal unit Gaussian white noise. Together, we arrive at an effective Langevin equation for $\varphi(t)$ whenever $\varphi$ lies outside the miscibility gap
\begin{align}
\label{eq:mdjasndqnqin}
  \partial_t \varphi = R_A(\varphi) + \sqrt{\frac{2\epsilon}{V} M_A(\varphi)}\, \Lambda(t) \,.
\end{align}

In contrast, for phase separated states where $\rho_{B,\eq}$ is inhomogeneous, we have
\begin{equation}
\label{eq:nionqidnqjssanddsjhdbqw}
\frac1V\int\dif\ve{x}\, R_A(\rho(\ve{x},t)) \approx\frac1V\int\dif\ve{x}\, R_A(\rho_{B,\eq}(\ve{x})) \approx \phi_+ R_A(\rho_+) + \phi_- R_A(\rho_-)\,,
\end{equation}
where we recall that the densities of the two bulk phases are $\rho_\pm = \rho_0\pm\sqrt{\alpha/\beta}$, their volumes are $\phi_\pm(\varphi) V$ obeying \eqref{eq:volumefracs}, and the volume of the interfacial region is negligible for $V$ large.
To find the stochastic term, we again use the additivity of Gaussian variances to obtain
\begin{align}
  \frac1V \int\dif\ve{x}\, \left(\sqrt{2\epsilon M_A(\rho_{B,\eq}(\ve{x},t))} \, \Lambda_A(\ve{x},t) \right) &= \sqrt{\frac{2\epsilon}{V^2} \left(\int\dif\ve{x}\, M_A(\rho_{B,\eq}(\ve{x},t))\right)} \,\Lambda(t)
\end{align}
where $\Lambda$ is a purely temporal unit Gaussian white noise. As before, approximating the profile as two homogenous bulk phases at densities $\rho_\pm$ separated by an interface of negligible width, we have 
\begin{align}
\label{eq:nionqidnqjssandjaksd}
  \frac{1}{V} \left(\int\dif\ve{x}\, M_A(\rho_{B,\eq}(\ve{x},t)\right) &\approx \phi_+ M_A(\rho_+)+ \phi_- M_A(\rho_-) \,.
\end{align}
This means that for phase-separated states the Langevin equation \eqref{eq:mdjasndqnqin} is replaced by
\begin{align}
\label{eq:mdjasndqnqin2}
  \partial_t \varphi = R_{\text{PS}}(\varphi) + \sqrt{\frac{2\epsilon}{V} M_{\text{PS}}(\varphi)}\, \Lambda(t) 
\end{align}
where we have used (\ref{eq:nionqidnqjssanddsjhdbqw},\ref{eq:nionqidnqjssandjaksd}) to define the global reaction rate and mobility of the phase-separated state:
\begin{align}
\label{eq:effparamPS}
  R_{\text{PS}}(\varphi) &:= \phi_+(\varphi) R_A(\rho_+) + \phi_-(\varphi) R_A(\rho_-)\,, \\
\label{eq:effparamPS2}
  M_{\text{PS}}(\varphi) &:= \phi_+(\varphi) M_A(\rho_+)+ \phi_-(\varphi) M_A(\rho_-)\,. 
\end{align}

Bringing everything together, we have derived the Langevin equation for the global density in the limit of slow reactions and weak noise
\begin{align}
\label{eq:langevin_effective}
  \partial_t \varphi = R(\varphi) + \sqrt{2\tilde{\epsilon} M(\varphi)} \Lambda(t)  
\end{align}
where $\tilde{\epsilon} := \epsilon/ V$ and recalling the definitions (\ref{eq:RA},\ref{eq:effparamPS})
\begin{align}
\label{eq:effparam}
  R(\varphi) &:= \begin{cases}
    R_A(\varphi) & \varphi\notin[\rho_-,\rho_+] \\
     R_{\text{PS}}(\varphi)& \varphi\in[\rho_-,\rho_+]  
  \end{cases},
  \\
  \label{eq:effparam2}
  M(\varphi) &:= \begin{cases}
    M_A(\varphi) & \varphi\notin[\rho_-,\rho_+] \\
   M_{\text{PS}}(\varphi) & \varphi\in[\rho_-,\rho_+]  
 \end{cases}.
\end{align}
We see that in the phase-separated regime the global reaction rate $R$ in \eqref{eq:langevin_effective} is a linear combination of the Model A reaction rates $R_A$ evaluated at the binodal densities $\rho_\pm$; similar remarks hold for the global mobility $M$. As anticipated above, the two binodal densities are the only parameters of the conservative  (Model B) dynamics that enter the Langevin equation \eqref{eq:langevin_effective}. The situation would be more complicated outside the slow reaction limit considered here.

\begin{figure}[bt]
\includegraphics[width=0.8\linewidth]{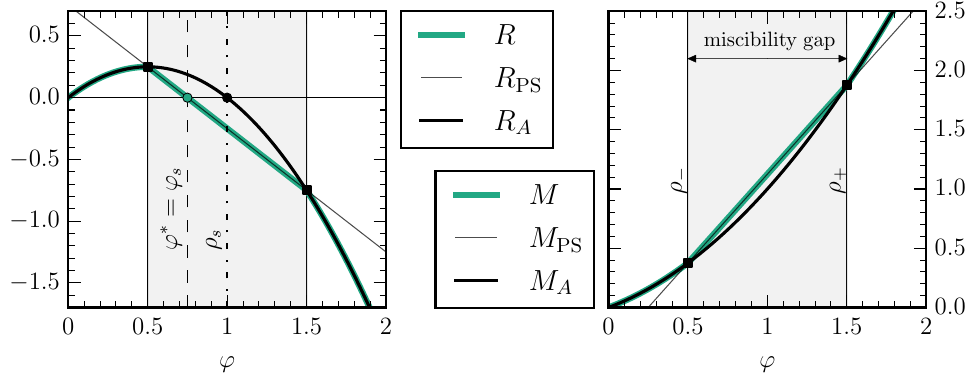}
\caption{Construction of the effective reaction rate $R$ and effective mobility $M$, starting from the logistic reaction rate $R_A(\rho)$ and mobility $M_A(\rho)$ for $\rho_- = 0.35, \rho_+ = 1.5$. To construct $R$, take $R_A$, then find $R_{\text{PS}}$ as the linear interpolation between $R_A(\rho_\pm)$ and finally set $R$ to $R_{\text{PS}}$ for $\rho\in[\rho_-,\rho_+]$ and $R_A$ else. The construction is analogous for $M$.  }
\label{fig:logistic_rate}
\end{figure}

The construction of $R$ and $M$ is illustrated in Fig.~\ref{fig:logistic_rate} for the logistic model. The solid black lines show the logistic reaction rate $R_A$ and mobility $M_A$, which apply outside the mobility gap, $\varphi \notin [\rho_-, \rho_+]$, while the grey lines show $R_{\text{PS}}$ and $M_{\text{PS}}$ which apply inside the gap only. The result for the global rate $R$ and mobility $M$ is found by stitching these functions together piecewise, giving (\ref{eq:effparam},\ref{eq:effparam2}), which are plotted as
the green lines in Fig.~\ref{fig:logistic_rate}.

\subsection{Effective free energy and mean time to extinction}

To obtain an effective free energy for $\varphi$ from the Langevin equation~\eqref{eq:langevin_effective} 
 we first define an effective chemical potential
\begin{align}
\label{eq:ndskajndjasda}
\mu(\varphi) &:= -\frac{R(\varphi)}{M(\varphi)}  = \begin{cases}
    \mu_A(\varphi) & \varphi\notin[\rho_-,\rho_+] \\
    \mu_{\text{PS}}(\varphi) & \varphi\in[\rho_-,\rho_+]  
 \end{cases}.
\end{align}
where $\mu_{\text{PS}} = -R_{\text{PS}} / M_{\text{PS}}$. Using (\ref{eq:effparamPS},\ref{eq:effparamPS2}) and assuming $R_A(\rho_-)\neq R_A(\rho_+)$ and $M_A(\rho_-)\neq M_A(\rho_+)$ one sees that $\mu_{\text{PS}}$ is a ratio of two linear functions, so we parameterise it as
\begin{align}
\label{eq:njdsandadjajdsa}
  \mu_{\text{PS}}(\varphi) &= \gamma\, \frac{\varphi - \varphi^*}{\varphi - \varphi^\circ}
\end{align}
with
\begin{align}
\label{eq:newparameters}
  \varphi^*  &:= \frac{R_A(\rho_+) \rho_- - R_A(\rho_-) \rho_+}{R_A(\rho_+) - R_A(\rho_-)}, \quad \varphi^\circ  := \frac{M_A(\rho_+) \rho_- - M_A(\rho_-) \rho_+}{M_A(\rho_+)  - M_A(\rho_-) }, 
  \quad \gamma := \frac{R_A(\rho_+) - R_A(\rho_-)}{M_A(\rho_+) - M_A(\rho_-)}.
\end{align}
The following results are straightforwardly extended to the degenerate cases  $R_A(\rho_-)= R_A(\rho_+)$ and $M_A(\rho_-)= M_A(\rho_+)$, we omit the details for brevity. Note that since the mobility $M$ cannot be negative, $\varphi^\circ$ always lies outside the gap.

The parameter $\varphi^*$ marks the (unique) zero of the reaction rate $R_{\text{PS}}$, which may lie inside or outside the miscibility gap.
In the latter case, the only fixed points of $R$ are those of $R_A$, 
 so the fixed points of the model-A dynamics carry over to the AB-type model unchanged.
However, if $\varphi^*$ lies within the miscibility gap, it marks the location of a new fixed point in $R$. This happens whenever $R_A(\rho_\pm)$ have opposite signs, and the new fixed point is stable if $R_A(\rho_+) < 0 < R_A(\rho_-)$, as illustrated in Fig.~\ref{fig:logistic_rate}.

In what follows, we consider the MTE for systems initialised at a stable fixed point of $R(\varphi)$, which might be either inside the miscibility gap (located at $\varphi=\varphi^*)$, or outside it (in which case it must be a stable fixed point of $R_A$).  We denote this stable fixed point by $\varphi_s$.
To find the MTE we 
integrate Eq.~\eqref{eq:ndskajndjasda} to find the effective free energy for the dynamics of the global density $\varphi(t)$:
\begin{align}
\label{eq:F}
  F(\varphi) = \int\dif\varphi\,\mu(\varphi) = \begin{cases}
    F_A(\varphi) - F_A(\rho_-) + F_{\text{PS}}(\rho_-) & \varphi \leq \rho_- \\
    F_{\text{PS}}(\varphi) &  \rho_- < \varphi \leq \rho_+ \\
    F_A(\varphi) - F_A(\rho_+) + F_{\text{PS}}(\rho_+) & \rho_+ < \varphi
 \end{cases}
\end{align}
where $F_\text{PS}$ is given by
\begin{equation}
\label{eq:F_PS}
  F_{\text{PS}}(\varphi) = \int\dif\varphi\, \mu_{\text{PS}}(\varphi) = \gamma\left( (\varphi- \varphi^*) - (\varphi^* - \varphi^\circ)\log\left|\frac{\varphi  - \varphi^\circ}{\varphi^* - \varphi^\circ}\right| \right) + C \,.
\end{equation}
where $C$ is a constant of integration, chosen such that $F(\varphi_s)=0$.  If $\varphi_s$ is inside the miscibility gap ($\varphi_s=\varphi^*$) then $C=0$.

Then the MTE is the mean first passage time to a state with vanishing density $\varphi=0$, starting from global density $\varphi_s$.  This is
\begin{equation}
\tau := \big\langle \inf\{t>0: \varphi(t) = 0\} \big| \varphi(0) = \varphi_s \big\rangle
\end{equation}
where the average is taken over trajectories of \eqref{eq:langevin_effective}, starting at $\varphi_s$.  We work in the small noise limit which is $\tilde\epsilon\to0$ in \eqref{eq:langevin_effective}.  For consistency with our previous analysis this should be achieved by taking $\epsilon\to0$ at fixed $V$.  (Taking $V\to\infty$ at fixed $\epsilon$ also yields a small-noise limit but its analysis requires a more careful analysis of the phase separated states, since \eqref{eq:approx} is only accurate for small $\epsilon$.)
For small $\tilde\epsilon$, the mean time to extinction $\tau$ then obeys asymptotically \cite{gardiner2009,doering2005}
\begin{align}
\label{eq:djsabdwqbdqdwq}
  \tilde \epsilon \log \tau \to F^* \qquad \hbox{with} \qquad F^*=  \sup\{ F(\varphi) :\varphi \in (0,\varphi_s)\}
\end{align}
where we used that $F(\varphi_s) = 0$. 
Physically, one sees that the MTE is controlled by the largest (effective) free energy barrier between $\varphi=\varphi_s$ and $\varphi=0$.
As a point of comparison for $F^*$ we also define 
\begin{equation}
\label{eq:FAstar}
F^*_A=  \sup\{ F_A(\varphi) :\varphi \in (0,\rho_s)\}
\end{equation}
which is the free energy barrier for a reference system where phase separation is absent (as may be achieved for example by assigning $\alpha$  in \eqref{eq:Bfree-dens} to some arbitrary negative value).  
The density $\rho_s$ appearing in \eqref{eq:FAstar} should be a finite-density fixed point of $R_A$,  chosen to ensure comparability with \eqref{eq:djsabdwqbdqdwq}.  For the logistic and Schl\"ogl models, the only possible choice is to take $\rho_s$ in \eqref{eq:FAstar} equal to the parameter $\rho_s$ that appears in their definitions\footnote{For a general model: If the stable fixed point $\varphi_s$ in \eqref{eq:djsabdwqbdqdwq} is outside the miscibility gap then we take $\rho_s=\varphi_s$ which is a stable fixed point of $R_A$; if $\varphi_s$  is inside the miscibility gap then there must also be at least one stable fixed point of $R_A$ in the gap, so a suitable choice for $\rho_s$ is always available.}.
Analogous to \eqref{eq:djsabdwqbdqdwq}, the MTE of the reference system scales as $\tilde \epsilon \log \tau_A \to F_A^*$, and the sign of
\begin{equation}
\Delta F^* = F^* - F_A^*
\end{equation}
determines whether extinction is speeded up by phase separation ($\Delta F^*<0$, MTE decreases), or slowed down ($\Delta F^*>0$, MTE increases).  The following subsections address this point for our chosen examples.

\subsection{Extinction with Phase Separation in the Logistic Model}
\label{sec:extinctionLM}
 
 \begin{figure}[bt]
\includegraphics[width=0.9\linewidth]{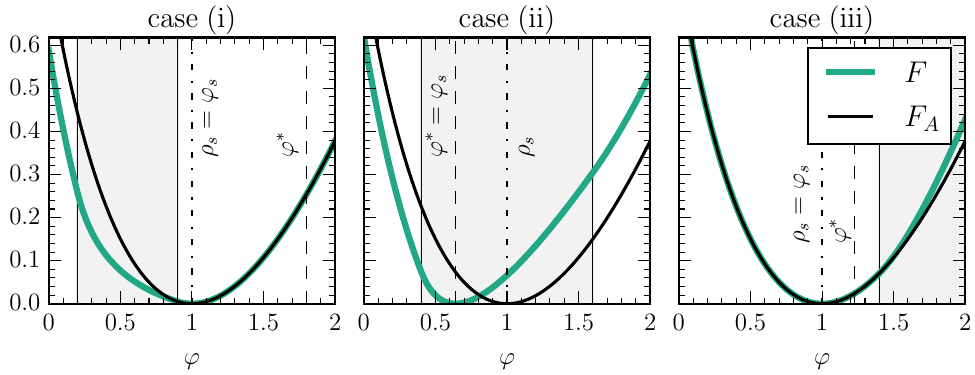}
\caption{Comparison between the effective free energy of a phase-separating logistic model $F$ with the free energy $F_A$ in the absence of phase separation. The fixed point of the effective reaction dynamics $\varphi^*$ is marked by the dashed line. Left to right: cases (i,ii) and (iii) as defined in the text.  Model parameters: $\rho_-/\rho_s = 1.2,0.4,0.2$, $\rho_+/\rho_s = 2.8,1.6, 0.9$.
}
\label{fig:logistic}
\end{figure}

For the logistic model, the parameters in \eqref{eq:newparameters} are
\begin{align}
\label{equ:logi-phi*-etc}
  \varphi^* = \frac{\rho_+\rho_-}{\rho_+ + \rho_- - \rho_s}, \quad \varphi^\circ = \frac{\rho_+\rho_-}{\rho_+ + \rho_- + \rho_s}, \quad \gamma = 2 \frac{\rho_s - \rho_+ - \rho_-}{\rho_s + \rho_+ + \rho_-}\,.
\end{align}
Fig.~\ref{fig:logistic} compares the corresponding effective free energy $F(\varphi)$ from \eqref{eq:F} with $F_A(\varphi)$ from \eqref{eq:dsandqwhudehfgiobjfv}. 
We distinguish the following three cases
\begin{center}
\begin{tikzpicture}

\node at (-4, 0) {case (i)};
\node at (-4, -0.6) {case (ii)};
\node at (-4, -1.2) {case (iii)};

\node at (0, -1.2) {$0 < \rho_s < \rho_- < \rho_+$};  
\node at (0, -0.6) {$0 < \rho_- < \rho_s < \rho_+$};  
\node at (0, 0) {$0 < \rho_- < \rho_+ < \rho_s$};  

\draw[line width=0.5pt] (4,-1.2) -- (8, -1.2) ;
\node[circle,fill=black,minimum size=5pt, inner sep=0pt] (c) at (4, -1.2){};
\node at (5, -1.2) {\Large $\times$};
\node at (6, -1.2) {$[$};
\node at (7, -1.2) {$]$};
\node at (7.89, -1.2) {$\blacktriangleright$};
\node at (4.5, -1.2) {$>$};
\node at (5.5, -1.2) {$<$};
\node at (6.5, -1.2) {$<$};
\node at (7.4, -1.2) {$<$};

\draw[line width=0.5pt] (4, -0.6) -- (8, -0.6) ;
\node[circle,fill=black,minimum size=5pt, inner sep=0pt] (c) at (4, -0.6){};
\node at (6, -0.6) {\Large $\times$};
\node at (5, -0.6) {$[$};
\node at (7, -0.6) {$]$};
\node at (7.89, -0.6) {$\blacktriangleright$};
\node at (4.5, -0.6) {$>$};
\node at (5.5, -0.6) {$>$};
\node at (6.5, -0.6) {$<$};
\node at (7.4, -0.6) {$<$};

\draw[line width=0.5pt] (4, 0) -- (8, 0) ;
\node[circle,fill=black,minimum size=5pt, inner sep=0pt] (c) at (4, 0){};
\node at (7, 0) {\Large $\times$};
\node at (5, 0) {$[$};
\node at (6, 0) {$]$};
\node at (7.89, 0) {$\blacktriangleright$};
\node at (4.5, 0) {$>$};
\node at (5.5, 0) {$>$};
\node at (6.5, 0) {$>$};
\node at (7.4, 0) {$<$};

\end{tikzpicture}
\end{center}
whose diagrammatic representation was introduced earlier, with the addition of $[$ and $]$ to  delineate the miscibility gap. It is easily verified that in all cases the effective free energy barriers to extinction [that is, the suprema in (\ref{eq:djsabdwqbdqdwq},\ref{eq:FAstar})] occur at $\varphi\to0$.  From \eqref{eq:dsandqwhudehfgiobjfv}, the barrier height without phase separation is linear in the carrying capacity $F_A(0) = 2(2\log2 - 1) \rho_s$.  We now treat the three cases in detail.

In case (i), the extinction pathway from a uniform initial state at $\rho_s$ passes through a phase separated state, and then becomes uniform again before the absorbing state is reached.  The illustrative example of this case in Fig.~\ref{fig:logistic} has $\Delta F^*<0$, so this phase separation tends to speed up extinction.  In fact, this case has always $\Delta F^*\leq0$: one sees from \eqref{eq:logistic} that $R_A''(\rho)<0$ and $M_A''(\rho)>0$.  This concave/convex structure means that the linear interpolations in (\ref{eq:effparamPS},\ref{eq:effparamPS2}) obey $R_{\rm PS}(\varphi) < R_A(\varphi)$ and $M_{\rm PS}(\varphi) > M_A(\varphi)$ within the miscibility gap.  Then \eqref{eq:ndskajndjasda} yields $\mu_{\rm PS}(\varphi) \geq \mu_{A}(\varphi)$ so recalling that the effective barrier to extinction is at $\varphi=0$ and that $\varphi_s=\rho_s$ in this case, one has 
\begin{equation}
 F^* = -\int_0^{\rho_s} \mu(\varphi') \dif\varphi' \leq  -\int_0^{\rho_s} \mu_A(\varphi') \dif\varphi' =  F_A^*
\label{equ:logi-deltaF*}
\end{equation}
 as required.

In case (ii), the stable fixed point $\varphi=\varphi_s$ is phase separated, and the path to extinction involves a reduction in the volume of the dense phase until the edge of the miscibility gap is reached, after which the system approaches extinction in a homogeneous state.  The illustrative example of this case in Fig.~\ref{fig:logistic} has $\Delta F^*<0$, and this result is also general.  To see this, note that $\mu<\mu_A$ as before, and that \eqref{equ:logi-phi*-etc} implies $\varphi^*<\rho_s$, and that $\mu_A(\varphi)<0$ for $\varphi<\rho_s$, so the analogue of \eqref{equ:logi-deltaF*} is $\Delta F^* = -\int_0^{\varphi_*} \mu(\varphi') \dif\varphi' \leq  -\int_0^{\rho_s} \mu_A(\varphi') \dif\varphi' = \Delta F_A^*$.

In case (iii) the path to extinction does not involve any phase separation, because the miscibility gap occurs at densities higher than the carrying capacity $\rho_s$.  Hence $F^* = F_A^*$ in this case and $\Delta F^*=0$ (an illustrative example is given in Fig.~\ref{fig:logistic}).  In general, we see that phase separation never slows down extinction in the logistic model, $\Delta F^*\leq 0$, which is attributable to the convex/concave properties of $R_A$ and $M_A$.

The effect of phase separation on extinction times in case (i) is an intrinsically non-equilibrium effect: If 
$F$ was an equilibrium free energy then it would be a state function, and the difference $F^*=F(0)-F(\varphi_s)$ between the free energies of two homogeneous states could never be affected by the presence or absence of a miscibility gap between $\varphi=0$ and $\varphi=\varphi_s$.  (The mean first passage time from $\varphi_s$ to zero could be affected by intermediate free energy barriers but these are absent in this case.)  Hence, while the noisy gradient flow \eqref{eq:langevin_effective} resembles an equilibrium model, the effective free energy that governs this flow is not a state function for the underlying field theory. Instead, $F$ must be constructed from the dynamics by integrating the effective chemical potential between initial and final states, which is affected by phase separation at intermediate densities.  This fact reinforces our terminology of \emph{effective} free energy and chemical potential for $F$ and $\mu$ respectively.

\subsection{Extinction with Phase Separation in the Schl\"ogl Model}
\label{sec:extinctionSM}

The concave/convex structure of $R_A$ and $M_A$ is a specific feature of the logistic model.  As a more generic example, we now consider the Schl\"ogl  model.  We focus on case (i) as defined above: the system is homogeneous at its stable fixed point $\varphi=\rho_s$ with a miscibility gap between this fixed point and the extinct state at $\varphi=0$.
Recall that the unstable fixed point of the Schl\"ogl model is at density $\rho_u<\rho_s$.  Hence we divide the original case (i) of the logistic model into four subcases, according to the location of $\rho_u$ relative to the miscibility gap:
\begin{center}
\begin{tikzpicture}

\node at (-4, 0) {case $(a)$};
\node at (-4, -0.6) {case $(b)$};
\node at (-4, -1.2) {case $(c)$};
\node at (-4, -1.8) {case $(d)$};

\node at (0, 0) {$0 < \rho_- < \rho_+ < \rho_u < \rho_s$};
\node at (0, -0.6) {$0 < \rho_- < \rho_u < \rho_+ < \rho_s$};
\node at (0, -1.2) {$0 < \rho_u < \rho_- < \rho_+ < \rho_s$};
\node at (0, -1.8) {$\rho_u < 0 < \rho_- < \rho_+ < \rho_s$};

\draw[line width=0.3pt] (4,0) -- (9, 0) ;
\node at (4, 0) {\Large $\times$};
\node at (5, 0) {$[$};
\node at (6, 0) {$]$};
\node[circle,fill=black,minimum size=5pt, inner sep=0pt] (c) at (7, 0){};
\node at (8, 0) {\Large $\times$};
\node at (9, 0) {$\blacktriangleright$};
\node at (4.5, 0) {$<$};
\node at (5.5, 0) {$<$};
\node at (6.5, 0) {$<$};
\node at (7.5, 0) {$>$};
\node at (8.5, 0) {$<$};

\draw[line width=0.3pt] (4, -1.2) -- (9, -1.2) ;
\node at (4, -1.2) {\Large $\times$};
\node[circle,fill=black,minimum size=5pt, inner sep=0pt] (c) at (5, -1.2){};
\node at (6, -1.2) {$[$};
\node at (7, -1.2) {$]$};
\node at (8, -1.2) {\Large $\times$};
\node at (9, -1.2) {$\blacktriangleright$};
\node at (4.5, -1.2) {$<$};
\node at (5.5, -1.2) {$>$};
\node at (6.5, -1.2) {$>$};
\node at (7.5, -1.2) {$>$};
\node at (8.5, -1.2) {$<$};

\draw[line width=0.3pt] (4, -0.6) -- (9, -0.6) ;
\node at (4, -0.6) {\Large $\times$};
\node at (5, -0.6) {$[$};
\node[circle,fill=black,minimum size=5pt, inner sep=0pt] (c) at (6, -0.6){};
\node at (7, -0.6) {$]$};
\node at (8, -0.6) {\Large $\times$};
\node at (9, -0.6) {$\blacktriangleright$};
\node at (4.5, -0.6) {$<$};
\node at (5.5, -0.6) {$<$};
\node at (6.5, -0.6) {$>$};
\node at (7.5, -0.6) {$>$};
\node at (8.5, -0.6) {$<$};

\draw[line width=0.3pt] (4, -1.8) -- (9, -1.8) ;
\node[circle,fill=black,minimum size=5pt, inner sep=0pt] (c) at (4, -1.8){};
\node at (5, -1.8) {$[$};

\node at (7, -1.8) {$]$};
\node at (8, -1.8) {\Large $\times$};
\node at (9, -1.8) {$\blacktriangleright$};
\node at (4.5, -1.8) {$>$};
\node at (5.6, -1.8) {$>$};
\node at (6.4, -1.8) {$>$};
\node at (7.5, -1.8) {$>$};
\node at (8.5, -1.8) {$<$};

\end{tikzpicture}
\end{center}
The effective free energy barrier [i.e., the supremum in \eqref{eq:djsabdwqbdqdwq}] is 
given by
\begin{align}
  F^* &=  \begin{cases}
    F(\varphi^*) &\text{ case $(b)$}  \\
    F(\rho_u) &\text{ cases $(a),(c)$} \\
    F(0) &\text{ case $(d)$}
 \end{cases}\,.
\end{align}
We discuss these cases in turn.

\begin{figure}[tb]
\includegraphics[width=\linewidth]{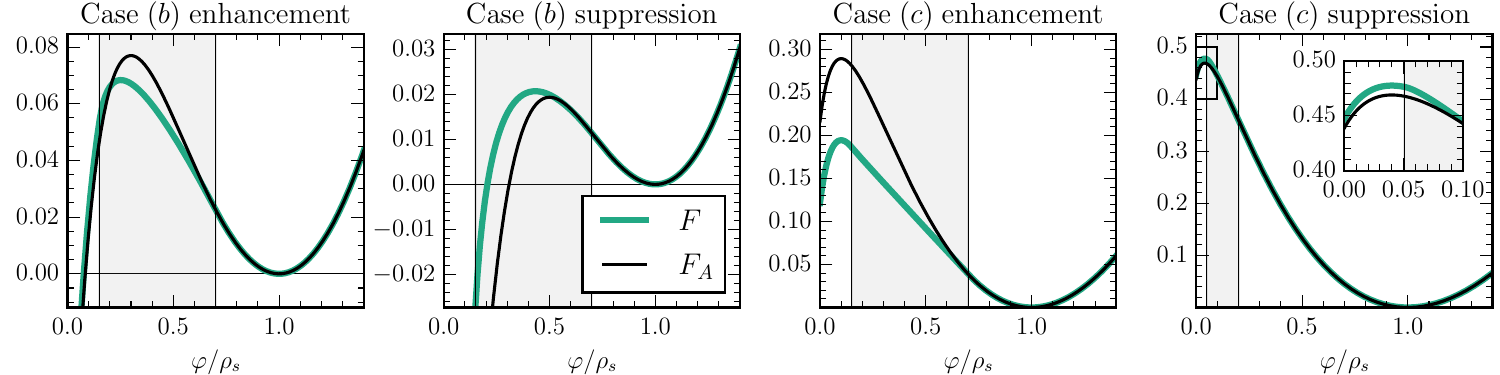}  
\caption{Extinction in the Schl\"ogl model. 
Free energies $F$ compared to $F_A$. Parameter values are $\rho_-/\rho_s = 0.15,0.15,0.15,0.0505$, $\rho_+/\rho_s = 0.7,0.7,0.7, 0.2$, $\rho_u/\rho_s = 0.3,0.5,0.1,0.04$. Note the different scales on the y-axis. 
}
\label{fig:schlogl1}
\end{figure}

For case $(a)$ the barrier lies at $\rho_u$, and phase separation occurs on the relaxational pathway from there to extinction: this does not affect the barrier, $F^*=F_A(\rho_u)$.  This is similar to case (iii) for the logistic model, and underlines that phase separated states only affect the MTE if they occur somewhere between the stable fixed point and the top of the barrier.

For case $(b)$, the barrier lies inside the miscibility gap, so the phase separation does affect the MTE.  The Schl\"ogl model does not have the concave/convex structure of $R_A$ and $M_A$ that was found in the logistic model, so the MTE can be either increased or decreased. Fig.~\ref{fig:schlogl1} exemplifies both suppression and enhancement.  

In case $(c)$, the miscibility gap lies between the initial uniform state and the top of the barrier, similar to case (i) for the logistic model.  Again, there is no concave/convex structure for $R_A$ and $M_A$ so the MTE can be either increased or decreased, with examples shown in Fig.~\ref{fig:schlogl1}.  In that figure, the examples where the MTE is suppressed have small values of $\Delta F^*$. We find in general that this suppression is weak, and occurs for a rather narrow range of parameters.  However, the weakness seems to be a specific feature of this model; we are not aware of any constraints that restrict this suppression to be weak.

Finally, case $(d)$ concerns the extension of the Schl\"ogl model to negative $\rho_u$, as discussed in Sec.~\ref{sec:CBDK}.  The resulting fixed-point structure is the same as case (i) for the logistic model, including that the barrier is at $\varphi=0$.  Numerical scans indicate that phase separation always lowers the barrier height in this case, similar to the corresponding situation for the logistic case.

As well as the cases $(a)$-$(d)$, we consider one additional possibility, which is a subcase of (iii) in the logistic model, for which both $\rho_s$ and $\rho_u$ lie inside the miscibility gap:
\begin{center}
\begin{tikzpicture}

\node at (-4, 0) {case $(e)$};

\node at (0, 0) {$0 < \rho_- < \rho_u < \rho_s < \rho_+$};

\draw[line width=0.3pt] (4,0) -- (9, 0) ;
\node at (4, 0) {\Large $\times$};
\node at (5, 0) {$[$};
\node at (8, 0) {$]$};
\node[circle,fill=black,minimum size=5pt, inner sep=0pt] (c) at (6, 0){};
\node at (7, 0) {\Large $\times$};
\node at (9, 0) {$\blacktriangleright$};
\node at (4.5, 0) {$<$};
\node at (5.5, 0) {$<$};
\node at (6.5, 0) {$>$};
\node at (7.5, 0) {$<$};
\node at (8.5, 0) {$<$};

\end{tikzpicture}
\end{center}
In this case there are no finite-density fixed points for the dynamics of the global density, and one sees from \eqref{eq:effparamPS} that $R(\varphi)<0$ except for $\varphi=0$ which has $R(0)=0$.  This means that the system always drifts on average towards extinction so the MTE is of order 1 and extinction is not a rare event.

\begin{figure}[tb]
\includegraphics[width=\linewidth]{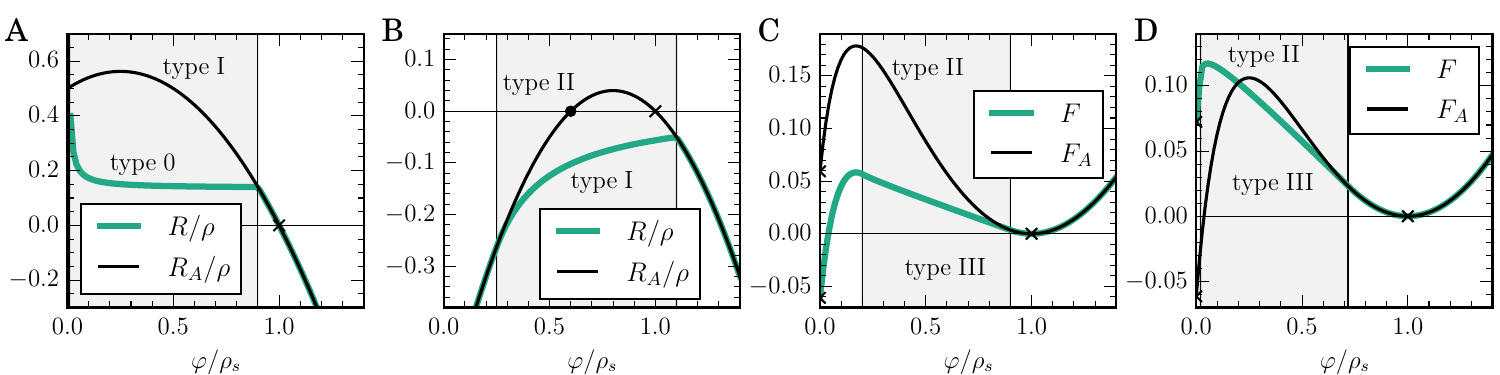}
\caption{Examples where phase separation changes the Allee type, compared to the homogeneous case
(recall Fig.~\ref{fig:allee}.)
\textbf{A}:~Behaviour of fitness function $R(\varphi)/\varphi$ showing change from type I in the homogeneous system to type 0 with phase separation, this is case (d) with parameters: $\rho_-/\rho_s = 0.01,\rho_+/\rho_s = 0.9,\rho_u/\rho_s = -0.5$;
\textbf{B}:~Fitness function showing change from type II to type I, this is case (e) with parameters: $\rho_-/\rho_s = 0.25,\rho_+/\rho_s = 1.1,\rho_u/\rho_s = 0.6$;
\textbf{C}:~Effective free energy, showing change from type II to Type III, this is case (c) with parameters: $\rho_-/\rho_s = 0.2,\rho_+/\rho_s = 0.9,\rho_u/\rho_s = 0.17$;
\textbf{D}:~Effective free energy, showing change from type III to type II, parameters: $\rho_-/\rho_s = 0.02,\rho_+/\rho_s = 0.717,\rho_u/\rho_s = 0.25$.
}
\label{fig:type_change}
\end{figure}

The examples of Fig~\ref{fig:schlogl1} show that the MTE can be either enhanced or suppressed by phase separation in the Schl\"ogl model.  We also find that the Allee type of the model can be changed by phase separation (recall the definitions of Sec.~\ref{sec:Alleetype}).  This phenomenon is illustrated in Fig.~\ref{fig:type_change} by several examples, as we now discuss.  Fig.~\ref{fig:type_change}A shows an example where the system without phase separation has a type I Allee effect, which is changed to type 0 (i.e., no Allee effect) on introducing phase separation.  The example comes from case (d) which has $R_A(\varphi)/\varphi$ positive as $\varphi\to0$, and its derivative is also positive there, consistent with type I.  A change to type 0 requires that $R(\varphi)/\varphi$ is monotonically decreasing between $\varphi=0$ and $\varphi=\varphi_s$, which requires in turn that the lower limit of the miscibility gap is at $\rho_-\to0$, as in the illustration.  To understand this change physically, recall that the maximum in $R_A$ comes from a crossover between co-operative effects at low density  and competition at high density.  In the presence of phase separation with $\rho_-\to0$, particles never exist in low-density states so the co-operative effects are irrelevant, and the (type I) Allee effect is lost.

Fig.~\ref{fig:type_change}B shows how a homogeneous system with Allee type II can be changed to type I by introducing phase separation.  This example comes from case (e) which means additionally that the stable fixed point is erased by the phase separation, and extinction ceases to be a rare event, as already explained above.  In other words, phase separation completely removes the effective free-energy barrier to extinction, which is generic for case (e).

Figs.~\ref{fig:type_change}(C,D) show examples where a phase separation changes the Allee type from type II to type III, and vice versa.  Recalling that the distinction between these types is based on whether the global minimum of $F(\varphi)$ is at $\varphi=0$ or $\varphi=\varphi_s$, this behaviour can occur naturally in cases where $F_A(\varphi_s)$ and $F_A(0)$ are close in the homogeneous system (with an intervening barrier).  The examples shown come from cases (c) and (b) respectively.  We note that the type changes shown in Figs.~\ref{fig:type_change}(A,C,D) require some parameter tuning, while that of Fig.~\ref{fig:type_change}B is generic throughout class (e).

The examples shown in Figs.~\ref{fig:type_change} illustrate all possible type changes.  For example, it is not possible for phase separation to change behaviour from type I to type II (or type III): this would require that $F$ has a local maximum that is not present in $F_A$, which means in turn that $R$ would have an additional unstable fixed point compared to $R_A$.  However, \eqref{eq:effparamPS} shows that $R_{\rm PS}(\varphi)$ is linear in $\varphi$ which means that the number of fixed points in the miscibility gap can never be increased by phase separation, hence local maxima cannot be created.  It is also impossible for phase separation to change the Allee type from 0 to I: Eqs.~(\ref{eq:effparamPS},\ref{eq:volumefracs}) imply 
\begin{equation}
\frac{\partial}{\partial\varphi}(R_{\rm PS}(\varphi)/\varphi) = \frac{R_A(\rho_+)\rho_--R_A(\rho_-)\rho_+}{\varphi^2(\rho_+-\rho_-)} \; ,
\end{equation} but if the homogeneous system has type 0 then $\frac{\partial}{\partial\rho}(R_A(\rho)/\rho)<0$ for all $\rho$, which implies that $R_A(\rho_+)\rho_- - R_A(\rho_-) \rho_+<0$.  Hence the derivative of $R_{\rm PS}(\varphi)/\varphi$ is negative and type I Allee effect is forbidden, even in the presence of phase separation.  This result also explains why phase separation cannot generate an Allee effect in the logistic model, since the homogeneous system has Allee type 0 in that case.

\subsection{Discussion of AB-type models}
\label{sec:discuss-AB}

We briefly summarise our analysis of these field-theoretic models, with a few additional remarks.  We have shown that the MTE may be suppressed or enhanced by adding fast bulk phase separation to reaction-diffusion systems.   Note also that while Fig.~\ref{fig:modelb} showed a one-dimensional illustration, our results apply independent of the spatial dimension $d$.  We also identified conditions on the convex/concave structure of $R_A$ and $M_A$ which ensure that the MTE can only be enhanced.  We showed (in the Schl\"ogl model) that the Allee type can be altered by phase separation.  We also observe that since $R$ has at most one fixed point in the miscibility gap, it is possible for phase separation to remove a pair of fixed points as in case (e) for the logistic model, or for multiple pairs of fixed points to be removed, in the general case.  [The parity (even/odd) of the number of fixed points in the gap remains always the same, it is fixed by the signs of $R_A(\rho_\pm)$.]

We recall that the type-AB  models are defined to have Gaussian noise and our detailed results are restricted to this case.  Calculation of the MTE for small-noise limits of reaction-diffusion systems requires analysis of Poissonian noise, but the qualitative behaviour that we find for our Gaussian models are mirrored in the Poissonian ones.  This is discussed in  Appendix \ref{sec:discrete_ab}.

Note also that the effective free energy $F(\varphi)$ computed in \eqref{eq:F} is related to the quasipotential for \eqref{eq:langevin_effective}.  Specifically, denote by $Q_{\varphi_s}$ the quasipotential computed for paths starting at the stable fixed point $\varphi_s$.  Then $F(\varphi)=Q_{\varphi_s}(\varphi)$ for $\varphi$ within the basin of $F$ that contains $\varphi_s$.  If $F$ has a barrier at finite $\varphi$ (separating the basin of attraction of $\varphi=\varphi_s$ from that of $\varphi=0$), then  $F\neq Q_{\varphi_s}$ for the basin containing the origin.  In this basin it is possible to identify $F$ with a suitable quasipotential $Q_{0}$, but the existence of extinct states complicates the interpretation of this function, and in any case the behaviour of $F$ in this regime is irrelevant for the MTE.

Finally, we note a similarity between our models that include phase separation, and situations where populations exist in two separate spatial locations (``islands'') as considered in~\cite{agranov2021}.  In that work, logistic growth takes place on each island, and the result is that a separated population increases the MTE, in contrast to our logistic model.  However, the calculations differ significantly, in diffusion/migration is fast in our Model AB but slow in~\cite{agranov2021}, and the noise structure is also different.  The resulting differences in behaviour highlight the broad range of phenomena that are supported in physical situations with separated populations.

%% file: sections/chapter2.tex
\newcommand{\GG}{\textrm{G}}

\section{Toy Model}
\label{sec:toy_model}

In the previous section, we analysed a reaction-diffusion system where chemical potential gradients drive the system towards  inhomogeneous states, via phase separation.  In this AB-type model the dynamics could be projected onto the total density, due to a separation of time scales.  The stochastic dynamics of this co-ordinate involves a weak noise, and we analysed rare fluctuations in a limit where this noise is small. We focussed on the extent to which inhomogeneous density fields affect the MTE.

To complement these results, we now present a very simple (``toy'') model where a few particles of two species hop between two lattice sites, as well as reacting with each other.  We analyse a similar limit to that of the previous Section, based on a joint limit of well-separated time scales and small noise.   We show how inhomogeneous mixing of the species can affect extinction events in this context too.
The results complement those of the previous section: the model is much simpler but it does capture the inherently discrete dynamics of reacting particles, and allows exact microscopic computations.
Furthermore, since the model does not involve projection onto a single reaction coordinate, it also allows us to explore how interactions of the type that promote phase separation change the {\em pathway} to extinction, not just the MTE.

When comparing the behaviour of the toy model with the field theory, note that the toy model has only two lattice sites, so there is no possibility for phase separation, but the hopping dynamics between the sites can lead to perfectly homogeneous mixing, or to inhomogeneous (imperfectly mixed) states.  It turns out that inhomogeneous mixing can only slow down the extinction, in contrast to the field-theoretic model, where it may be accelerated or slowed down.  We discuss these (and other) similarities and differences between the two models.

\subsection{Model}
\label{sec:toy model}

The toy model features reaction and diffusion, as in the AB-type model.
It is defined in a discrete state space with a limited number of particles distributed over $M$ lattice sites (we will later set $M=2$, which is the simplest non-trivial case).  
There are two types of particles, G(reen) and W(hite), which hop between the lattice sites. A chemical reaction, $2G \rightleftharpoons W$ leads to interconversion between the particle species.  Let $g_m$ be the number of $G$ particles on site $m$, and similarly $w_m$ for the number of $W$ particles.

There are on-site interactions between particles of different species.  As in the AB-type model, the reaction and hopping dynamics are governed by different interactions.  We write effective interaction energies on site $m$ as
\begin{equation}
E_R(g_m,w_m) = J_R g_m w_m , \qquad E_H(g_m,w_m) = J_H g_m w_m 
\end{equation}
which are associated with reactions and hopping respectively, where $J_H,J_R$ are coupling constants, defined so that positive values represent repulsive interactions between the species.  
We specify the particle dynamics in detail below, but first note that the reaction dynamics is defined to obey detailed balance with respect to the (non-normalised) grand-canonical distribution
\begin{equation}
\pi_R(g_1,w_1,g_2,w_2,\ldots,) = \prod_{i=1}^M \frac{1}{g_m ! w_m !} \exp\left[ \beta \mu_g g_m + \beta \mu_w w_m - \beta E_R(g_m,w_m) \right] \ ,
\label{equ:pi-gc}
\end{equation}
where $\mu_g,\mu_w$ are chemical potentials for the two species, and $\beta$ is the inverse temperature.  The hopping dynamics obeys detailed balance with respect to an analogous distribution $\pi_H$, which has the same definition except that $E_R$ is replaced by $E_H$.
For the special case $J_R=J_H$, the whole system respects detailed balance with respect to $\pi_R$.  Otherwise it has a non-equilibrium steady state.

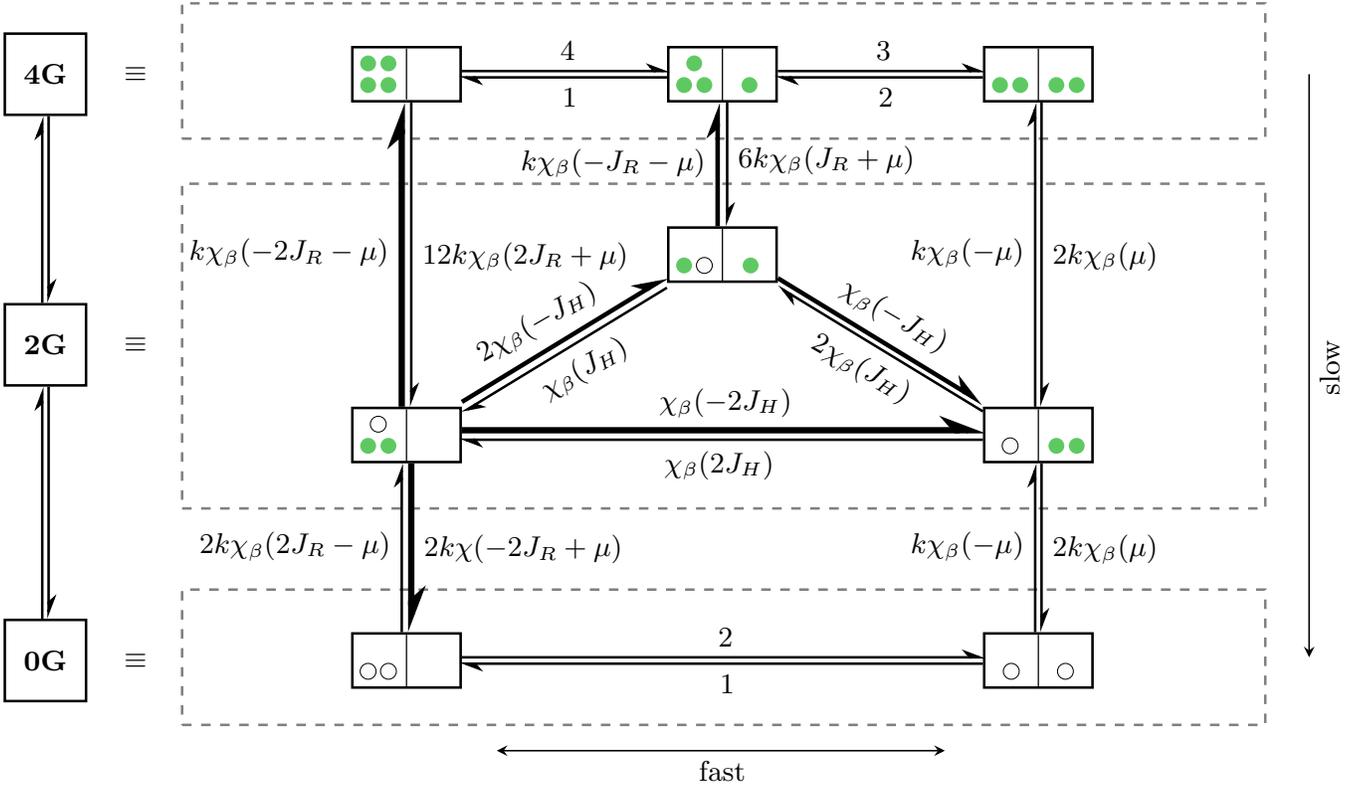
\begin{figure}[tb]
\centering
\input{sections/bigtikz.tex}
\caption{State space of the toy model in its 8-state representation.  The boxes represent microstates, where pairs of states that are related by interchanging sites have been grouped together (only one such state is depicted in each case).  Arrows represent the transitions between microstates, labelled by the rates $r$. The left column shows the coarse-grained states 0G, 2G and 4G and their relation to the microstates.  Thick arrows indicate transitions which are enhanced by positive couplings $J_R,J_H$, which we recall correspond to on-site inter-species repulsive interactions.  In the limit of slow reactions ($k\to0$) vertical transitions are of order $k$ (slow), while horizontal transitions are of order $1$ (fast). }
\label{fig:micro_statespace}
\end{figure}

The reaction dynamics consists of the transitions $(g_m, w_m) \to (g_m \pm 2, w_m \mp 1)$.  
Denoting transitions between microstates $x,y$ as $r[x\to y]$, we take
\begin{subequations}
\label{eq:mlmsdainwuqbdd}
\begin{align}
  r[(g_m,w_m) \to (g_m-2,w_m+1)] &= k g_m(g_m-1) \chi_\beta(\mu+J_R(g_m-2w_m-2)) \ ,  
  \\
  r[(g_m-2,w_m+1)\to (g_m,w_m)] &=  k (w_m + 1) \chi_\beta(-\mu-J_R(g_m-2w_m-2)) \ , 
\end{align}
\end{subequations}
where $\mu = 2\mu_g - \mu_w$ is the difference in chemical potential between reactants and products, the parameter $k$ sets the time scale for the reactions, and
 $\chi_\beta(x) = 1/(1+{\rm e}^{\beta x})$ is a function inspired by Glauber rates.
 Note that detailed balance with respect to $\pi_R$ requires $\chi_\beta(x) = {\rm e}^{-\beta x} \chi_\beta(-x)$, but this leaves considerable freedom in the choice of $\chi_\beta$;  alternative choices will be discussed briefly in Sec.~\ref{sec:sqrt}.

In a similar way, hopping of $G$ and $W$ between nearest neighbour sites $m,l$ has rates
\begin{subequations}
\label{eq:mlmsdainwuqbdd2}
\begin{align}
  r[(g_m+1,g_l) \to (g_m,g_l+1)] &=  (g_m+1) \chi_\beta(J_H (w_l-w_m)) \ , 
  \\
  r[(w_m+1,w_l) \to (w_m,w_l+1)] &= (w_m+1) \chi_\beta(J_H(g_l-g_m)) \ , 
\end{align}
\end{subequations}
which respect detailed balance with respect to $\pi_H$.  For $J_H=0$, this is free diffusion (the factors of $g_m+1$ and $w_m+1$ enter because each particle on site $m$ has an equal rate of hopping to site $l$, and the transition rate $r$ is the sum of these rates).

The quantity $n=\sum_m (g_m+2w_m)$ is conserved under the full dynamics.  To simplify the model as far as possible, we suppose that the system is prepared with $n=4$, and we consider $M=2$ lattice sites.  Then the full reaction dynamics reduces to a Markov chain with 14 microstates, which are labelled by the variables $(g_1,w_1,g_2,w_2)$.  Each state is conveniently denoted by a pair of boxes that represent the lattice sites, with their constituent boxes.  Moreover, the system is symmetric under interchange of lattice sites, which allows some pairs of microstates to be grouped together, for example the two microstates $(4,0,0,0)$ and $(0,0,4,0)$ containing 4 green particles on either lattice site and no white particles, represented pictorially as
\begin{center} 
\begin{tikzpicture}

\definecolor{colorA}{rgb}{0.369214, 0.788888, 0.382914}
\definecolor{colorB}{rgb}{0.99, 0.99, 0.99}

\node[rectangle,draw,minimum width=1.2cm,minimum height=0.6cm] (1) at (0,0) {};
\draw[thin] (0,-0.28) -- (0,0.32) ;
\node[rectangle,draw,,minimum width=1.2cm,minimum height=0.6cm] (6) at (2,0) {};
\draw[thin] (2,0-0.3) -- (2,0+0.32) ;
\node[circle,fill=colorA,minimum size=5pt, inner sep=0pt] (c) at (-0.425,-0.12){};
\node[circle,fill=colorA,minimum size=5pt, inner sep=0pt] (c) at (-0.2,-0.12){};
\node[circle,fill=colorA,minimum size=5pt, inner sep=0pt] (c) at (-0.425,0.12){};
\node[circle,fill=colorA,minimum size=5pt, inner sep=0pt] (c) at (-0.2,0.12){};
\node[circle,fill=colorA,minimum size=5pt, inner sep=0pt] (c) at (2+0.425,0-0.12){};
\node[circle,fill=colorA,minimum size=5pt, inner sep=0pt] (c) at (2+0.2,0-0.12){};
\node[circle,fill=colorA,minimum size=5pt, inner sep=0pt] (c) at (2+0.2,0+0.12){};
\node[circle,fill=colorA,minimum size=5pt, inner sep=0pt] (c) at (2+0.425,0+0.12){};
\end{tikzpicture} \; 
\end{center}
This means that the system can be modelled as an 8-state Markov chain, whose states are shown in Fig.~\ref{fig:micro_statespace}, together with the relevant transition rates $r$ as defined for the full 14-state model in (\ref{eq:mlmsdainwuqbdd},\ref{eq:mlmsdainwuqbdd2}).
We identify three subspaces according to the total number of $G$ particles, as indicated on the left part of the Figure.  
Notice that $J_R$ affects transitions between the subspaces, while $J_H$ only affects transitions within the 2G subspace  (which is the only subspace in which both species exist together).

\subsection{Timescale separation, reduction to reaction coordinate} 
\label{sec:projection2}

In Sec.~\ref{sec:projection1} we analysed the AB-type model by projecting the dynamical equation \eqref{eq:modelab} onto the mean density using a timescale separation between the reaction and diffusion dynamics. We now derive an equivalent projection for the toy model. 
The timescale separation occurs when $k \ll 1$, and this condition is assumed throughout the analysis.  Fig.~\ref{fig:micro_statespace} shows that all hopping (horizontal) transitions have fast rates (of order 1), while reaction (vertical) transitions are slow (rates are of order $k$). 
This leads to quasi-equilibrium within the subspaces 0G, 2G, and 4G, which we refer to as meso-states (recall that these are labelled by the number of $G$ particles).

We denote microstates of the system by $i,j,\dots$ and the meso-states as $I,J,\dots$.  For $k\ll 1$, the dynamics of the system is that a system started in state $i\in I$ relaxes quickly into a quasi-equilibrium distribution $\pi^I(i)$ which is positive only for $i\in I$ and normalised such that $\sum_{i\in I} \pi^I(i)=1$.  For large times $t$ (of order $k^{-1}$), the rest of the dynamics becomes asymptotically Markovian, with slow transitions between the mesostates, whose rates are
\begin{equation}
\label{eq:rmeso}
r_{\rm meso}(I\to J) = \sum_{i\in I, j\in J} \pi^I(i)  r(i \to j)  \; .
\end{equation}
For finite $k$, projection onto the mesostates introduces memory in the form of non-homogeneous rates \cite{lapolla2019}, but these are negligible as $k\to0$, when time-scale separation holds.
In this limit, the steady state of the system also factorises as \cite{gaveau2006}
\begin{align}
  \lim_{k\to0} \pi(i) = \pi^{I_i}(i)  \nu(I_i)  
\end{align}
where $I_i$ is the mesostate containing microstate $i$, and $\nu(I)$ is the  steady state probability of meso-state $I$ (asymptotically as $k\to0$).
For the toy model, the $\pi^I$ have the following distributions
\begin{align}
  \pi^{\text{4G}}(g_1,w_1,g_2,w_2) &=  \frac1{16} \frac{4!}{g_1!g_2!} \delta(g_1+g_2,4) \delta(w_1+w_2,0)
  \nonumber \\
  \pi^{\text{2G}}(g_1,w_1,g_2,w_2) &=  \frac12 \frac{2!}{g_1!g_2!} 
  [ p^{g_1}(1-p)^{g_2} \delta(w_1,1) + p^{g_2}(1-p)^{g_1} \delta(w_2,1)  ] \delta(g_1+g_2,2) \delta(w_1+w_2,1)
  \label{equ:pi-xG}
  \\ \nonumber
  \pi^{\text{0G}}(g_1,w_1,g_2,w_2) &=  \frac14 \frac{2!}{w_1!w_2!}  \delta(w_1+w_2,2)\delta(g_1+g_2,0)
\end{align}
where $p = (1+e^{\beta J_H})^{-1}$ and $\delta(x,y)$ is 1 for $x=y$ and 0 otherwise.
Here, $\pi^{\text{4G}}$ and $\pi^{\text{0G}}$ are Binomial distributions with parameter $1/2$, which correspond to homogeneous mixing of the particle (that is, the relative occupancies of the states are in accordance with their entropies).
Physically, $\pi^{\text{2G}}$ distributes the two G particles conditionally according to the position of the W particle:  the conditional distribution for $g_1$ is binomial with a parameter that depends on $w_1$.
This corresponds to inhomogeneous mixing in the $2\GG$ states, due to the interactions.  The only exception is $J_H=0$ where mixing is homogeneous: we use this case as  a point of comparison when analysing effects of inhomogeneity.

Now observe that since the three meso-states form a linear chain, their dynamics must obey detailed balance with respect to $\nu$, that is
\begin{equation}
\frac{\nu(2G)}{\nu(4G)} = \frac{ r_{\rm meso}(4G\to 2G) } { r_{\rm meso}(2G\to 4G) }  , \qquad
\frac{\nu(0G)}{\nu(2G)} = \frac{ r_{\rm meso}(2G\to 0G) } { r_{\rm meso}(0G\to 2G) } \ .
\label{equ:toy-meso-db}
\end{equation}

In the following, we study rare events that occur when time scales are well-separated ($k\ll1)$ and the temperature $T=1/\beta$ is small. The low-temperature limit plays the equivalent role of the low noise limit in Sec \ref{sec:projection1}.  Similar to that case, we assume that time scales are well-separated even as the noise strength tends to zero. In particular, we assume that $k$ is always much smaller than quantities of the form ${\rm e}^{-\beta J_R}$, even as we take the low-temperature limit.  To analyse this limit, we define an effective free energy relative to the 4G state:
\begin{align}
\label{eq:fnjnojwqhiwqeewq}
  f(I) = -\lim_{\beta\to\infty} \beta^{-1} \log \frac{\nu(I)}{\nu(4G)} 
\end{align}
corresponding to the effective free energy $F$ in \eqref{eq:F} for the AB-type model.

\subsection{Mean first-passage time to extinction}

We now discuss the analogue of an extinction transition in this model. 
For a system starting in the 4G meso-state, we analyse the probability that the system makes an excursion to 0G.  This is assumed to result in extinction by a mechanism 
similar to that described in Sec.~\ref{sec:CBDK}: we supplement the model by an absorbing (extinct) state $X$, such that the two $W$ particles that exist in the $0\GG$ state may disappear, leaving the system in an ``extinct'' state $X$ that has no particles at all:
\begin{center}
\begin{tikzpicture}[-,>=stealth,shorten >=1pt,auto,scale=1,node distance=3cm,thick,main node/.style={circle,draw,font=\Large}]

\definecolor{colorA}{rgb}{0.369214, 0.788888, 0.382914}
\definecolor{colorB}{rgb}{0.99, 0.99, 0.99}

\node[rectangle,draw,minimum width=0.9cm,minimum height=0.9cm] (X)  at (-2,0) {{X}};
\node[rectangle,draw,minimum width=0.9cm,minimum height=0.9cm] (0A)  at (0,0) {{0G}};
\node[rectangle,draw,minimum width=0.9cm,minimum height=0.9cm] (2A)  at (2,0) {{2G}};
\node[rectangle,draw,minimum width=0.9cm,minimum height=0.9cm] (4A)  at (4,0) {{4G}};

\draw[-{Stealth[left]}] ([yshift=-1pt,xshift=0pt]4A.west) -- ([yshift=-1pt,xshift=-1pt]2A.east) node[midway,left] {}; 
\draw[-{Stealth[left]}] ([yshift=1pt,xshift=0pt]2A.east) -- ([yshift=1pt,xshift=1pt]4A.west) node[midway,right] {};
\draw[-{Stealth[left]}] ([yshift=-1pt,xshift=0pt]2A.west) -- ([yshift=-1pt,xshift=-1pt]0A.east) node[midway,left] {}; 
\draw[-{Stealth[left]}] ([yshift=1pt,xshift=0pt]0A.east) -- ([yshift=1pt,xshift=1pt]2A.west) node[midway,right] {}; 

\draw[-{Stealth}] ([yshift=0pt,xshift=0pt]0A.west) -- ([yshift=0pt,xshift=-1pt]X.east) node[midway,right] {}; 

\end{tikzpicture}
\end{center}

Separation of this final extinction step of the dynamics allows us to treat the remaining processes as reversible (albeit without detailed balance) so that invariant measure can be defined.
It is convenient to restrict to situations where the $4\GG$ state is a local minimum of $f$, that is $f(4\GG)<f(2\GG)$.  We also assume that the steps from $4\GG$ to $0\GG$ are rate-limiting for the extinction process, which will be valid as long as  $r(\text{0G} \to \text{X})$ is not too small.
In this situation, we show in App.~\ref{sec:mte_calc} that
the mean first-passage time to extinction $\tau_{\rm ext}$ has low-temperature asymptotic behaviour
\begin{equation}
\beta^{-1} \log (k\tau_{\rm ext}) \to f^* 
\label{equ:tau-f*}
\end{equation}
with 
\begin{equation}
f^*={\rm max}[ f(2\GG),f(0\GG) ].
\label{equ:def-f*}
\end{equation}  
This is the physically-expected result that the extinction time is controlled by the largest barrier in $f$ between $4\GG$ and $0\GG$.
If this maximum is achieved by $2\GG$ then we interpret this as a form of Allee effect (either type II or III); if it is achieved by $0\GG$ then there is no meaningful Allee effect.  It is not possible to distinguish Allee types 0 and I in this model, since that distinction uses the slope of the reaction rate, so we group these these types into a composite ``type 0/I'' in this section.
On the other hand, we can distinguish type II and III based on the sign of $f(\text{0G})$. Following Sec.~\ref{sec:Alleetype} we arrive at the following interpretation of Allee types for the toy model 
\begin{align}
\label{eq:toy_allee_classification}
\begin{split}
  0 < f(\text{2G}) \leq f(\text{0G}) \qquad & \equiv\qquad \text{type 0/I (no Allee effect)}
  \\
  0  \leq f(\text{0G}) < f(\text{2G}) \qquad & \equiv\qquad\text{type II}
  \\
  f(\text{0G}) < 0 < f(\text{2G}) \qquad & \equiv\qquad\text{type III}
\end{split}
\end{align}
which is only based on the free energy. 
Recall that $f(4\GG)=0$ by definition and that we are taking $k\to0$ at the same time as our low-temperature limit, so that we maintain the separation of time scales between reactions and hopping.  

In the following we analyse the extinction time by computing $f^*$, focussing in particular on situations where it is significantly affected by inhomogeneous mixing within the meso-states ($J_H\neq0$).

\subsection{Effective free energies and dominant pathways between meso-states}

The effective free energy $f$ can be computed in two different ways: either by using \eqref{equ:toy-meso-db} to compute the invariant measure $\nu$ as a function of $\beta$ and then taking $\beta\to\infty$; or by identifying the large-$\beta$ behaviour of the rates $r_{\rm meso}$ in \eqref{equ:toy-meso-db} and using relations like
\begin{equation}
f(2\GG) = \lim_{\beta\to\infty} \beta^{-1} \log [r_{\rm meso}(2\GG\to4\GG)/k] - \lim_{\beta\to\infty} \beta^{-1} \log [r_{\rm meso}(4\GG\to2\GG)/k] .
\end{equation}
which follows from (\ref{equ:toy-meso-db},\ref{eq:fnjnojwqhiwqeewq}) since both limits exist.  
This second route is both convenient and physically-informative because the mesoscopic rates in \eqref{eq:rmeso} simplify in the low-temperature limit as
\begin{equation}
 \lim_{\beta\to\infty} \beta^{-1} \log r_{\rm meso}(I\to J) =  \sup_{i\in I, j\in J} \left( \lim_{\beta\to\infty} \beta^{-1} \log[ \pi^I(i)  r(i \to j)] \right)
 \label{equ:r-sup}
\end{equation}
Physically, the supremum picks out which of the microscopic rates dominates the transition between the meso-states, at low temperatures.  (There may be a single dominant rate or several degenerate ones.)  This leads to a decomposition of the parameter space of the model into different regimes, according to which are the dominant microscopic rates.

\begin{figure}
\includegraphics[width=0.8\linewidth]{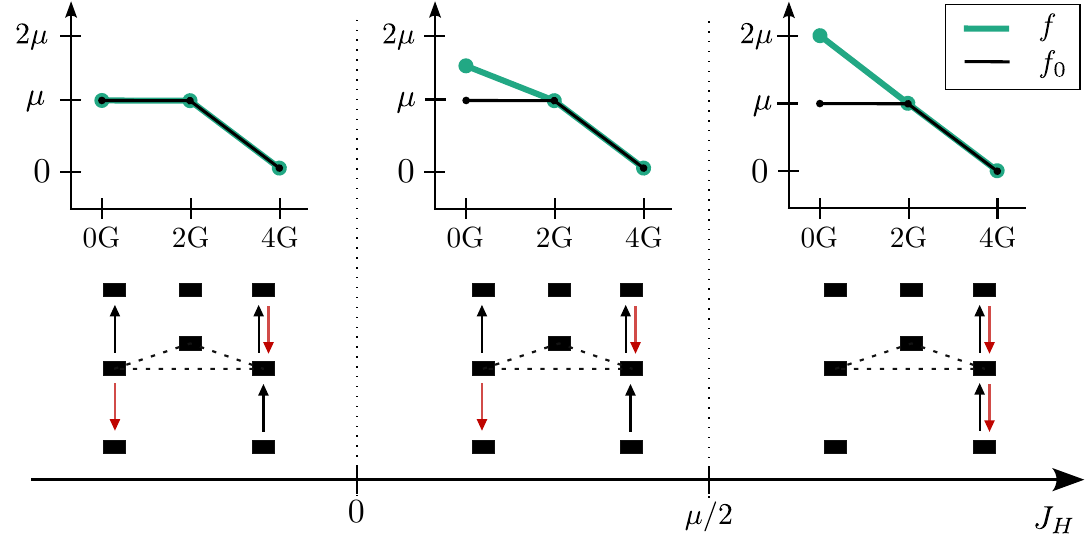}
\caption{
Behaviour of the toy model as a function of $J_H$, for the illustrative parameter regime of $0 < \mu < 2J_R$.  The upper panels show the form of the free energy $f$, compared with its form $f_0$ in the case $J_H=0$ (homogeneous mixing).
Arrows in the lower panels indicate the rates that dominate the suprema in \eqref{equ:r-sup}, where the layout of the microstates follows Fig.~\ref{fig:micro_statespace} (hence the 4G subspace is at the top and 0G at the bottom).  Red arrows indicate the dominant path to extinction (decreasing the number $G$ particles) while black arrow indicate transitions with increasing $G$.  Cases where multiple black arrows connect two subspaces occur when the maximiser of \eqref{equ:r-sup} is degenerate.  Dashed lines are reminders that transitions along the solid arrows are interspersed by many (fast) transitions within the $2\GG$ subspace.
}
\label{fig:ch2-illust}
\end{figure}

\subsubsection{Illustrative case: $0<\mu<2J_R$}

Depending on which microscopic rates dominate in \eqref{equ:r-sup}, we distinguish different regimes of behaviour.
To illustrate the generic phenomena supported by the model, we first suppose that $0<\mu<2J_R$ and consider the behaviour as a function of $J_H$.
Other values of $\mu,J_R$ will be discussed below.
We find as $\beta\to\infty$ that
\begin{align}
\beta^{-1} \log [r_{\rm meso}(4\GG\to 2\GG)/k] & \to -\mu
\\ 
\beta^{-1} \log [r_{\rm meso}(2\GG\to 4\GG)/k] & \to 0
\\ 
\beta^{-1} \log [r_{\rm meso}(0\GG\to 2\GG)/k] & \to 0
\end{align}
and 
\begin{equation}
\beta^{-1} \log [r_{\rm meso}(2\GG\to 0\GG)/k]  \to
 \begin{cases} 0 , & J_H\leq0
\\ -2J_H , & 0< J_H \leq \mu/2
\\ -\mu , & \mu/2 < J_H
 \end{cases}
\end{equation}
The microstate transitions that dominate these rates are illustrated in Fig.~\ref{fig:ch2-illust}.  It follows that $f(2\GG) = \mu$ and
\begin{equation}
f(0\GG) = \begin{cases} \mu , & J_H\leq0 \\ \mu+2J_H , & 0<J_H\leq \mu/2 \\ 2\mu , & \mu/2<J_H \,. \end{cases}
\end{equation}
Moreover we see that $f^*=f(0\GG)$ determines the MTE. 
This is a piecewise-linear function of $J_H$ and we see that moving away from the homogeneous case ($J_H=0$) can only increase $f^*$, which increases the MTE, slowing down extinction.
Another important case is $J_H=J_R$ so that the model is in equilibrium: this again slows extinction, compared to homogeneous mixing.
There is no Allee effect for these parameters.   

One may also apply transition path theory~\cite{metzner2009} to the mesoscopic model, to understand the dominant pathways to extinction.  At finite temperature, this requires computation of forward and backward committors from $2\GG$ to $0\GG$ and $4\GG$.  However, in the low-temperature limit considered here, the dominant transition pathways can be obtained from \eqref{equ:r-sup}, and are shown in Fig.~\ref{fig:ch2-illust}.  For example, if $J_H<0$ then trajectories to the extinct state overwhelmingly leave the $4\GG$ through the ``top right'' state [$(g_1,g_2)=(2,2)$] and enter $0\GG$ via the ``bottom-left'' state [$(w_1,w_2)=(2,0)$].
However, the separation of time scales in the model mean that trajectories leading to extinction have a non-trivial character: they move between the meso-states following the dominant pathways in Fig.~\ref{fig:ch2-illust}, but the fast hopping means that system equilibrates locally within $2\GG$, before transitioning to $0\GG$.  That is, a typical path to extinction makes multiple transitions along the dashed paths in Fig.~\ref{fig:ch2-illust} during the intermediate period between leaving $4\GG$ and arriving in $0\GG$.

Note also that for equilibrium systems, the dominant pathways between mesostates must always follow the same paths for both forward and backward transitions: this is the case for $J_H>\mu/2$ in Fig.~\ref{fig:ch2-illust} (which includes the equilibrium case $J_H=J_R$) but it does not hold in general.  Hence, the non-equilibrium nature of the microscopic model can strongly affect extinction pathways, even if the dynamics among meso-states have a time-reversible (detailed balanced) dynamics, as encoded by \eqref{equ:toy-meso-db}.

These observations help to rationalise the the effect of inhomogeneous mixing on extinction.  Recall that $J_H=0$ is the homogeneously mixed system and we analyse the other cases in comparison to this.  We find that $J_H<0$ has no effect on the extinction time, while $J_H>0$ slows down extinction.    One way to understand this slowing-down is that homogeneous mixing within the $2\GG$ meso-state means that all of the $\pi^{2\GG}(i)$ factors in \eqref{equ:r-sup} are of order unity, but if $J_H\neq0$ then some of the $\pi^{2\GG}(i)$ factors are suppressed by factors of order ${\rm e}^{\beta |J_H|}$ or
${\rm e}^{2\beta |J_H|}$, recall \eqref{equ:pi-xG}. 
Whenever the dominant path from 2G to 0G passes through an exponentially suppressed state $i$ in 2G, the rate $r_{\rm meso}(2\GG\to0\GG)$ is suppressed, which tends to increase the barrier height $f(\text{0G})$.  However, if
the occupation of $i$ in 2G becomes too low, the dominant extinction transition changes to avoid the suppressed state $i$, by routing through a different state in 2G. This effect can be seen in Fig.~\ref{fig:ch2-illust}:  For $J_H < \mu/2$ the dominant pathway from 4G to rate 0G (marked in red) takes the ``top right" transition into 2G and the ``bottom left" transition into 0G.  For $0<J_H<\mu/2$, the $r_{\rm meso}(2\GG\to0\GG)$ is suppressed by the $\pi^{2\GG}(i)$ factor for ``leftmost'' state in 2G.  However, for $J_H > \mu/2$ the occupation of the this becomes so low that the dominant pathway from 2G to 0G switches to the bottom right. 
All together, this implies that inhomogeneous mixing always tends to increase $f^*$ and hence slow down extinction in this regime.

\begin{figure*}
\includegraphics[width=\linewidth]{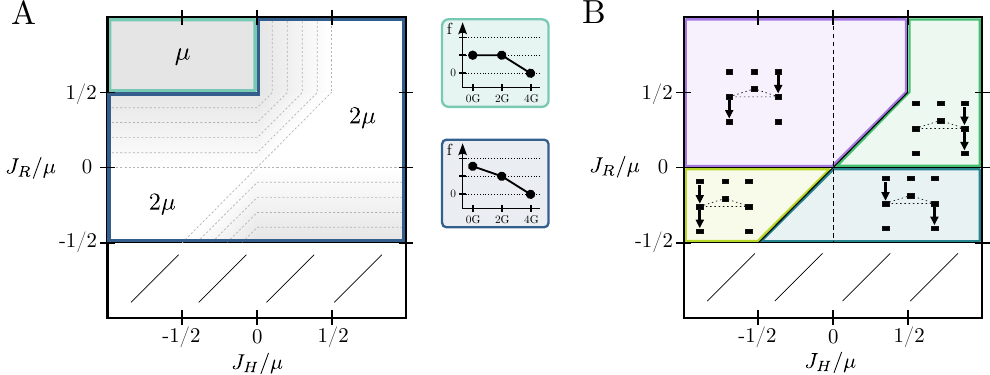}
\caption{
Effective free energy and dominant pathways between meso-states for $\mu>0$ as a function of general $J_R$ and $J_H$. 
The top regions $J_R > \mu/2$ in both panels corresponds to the illustrative case shown in  Fig.~\ref{fig:ch2-illust}.
\textbf{A}: The free energy barrier $f^* = f(\text{0G})$. Note the piecewise linear interpolation between the values $\mu$ and $2\mu$. Side panels show qualitative shapes of the free energy in the marked regions. 
\textbf{B}: The four distinct dominant pathways to extinction.  The representation similar to Fig.~\ref{fig:ch2-illust}, the rectangles represent micro-states, and the arrows depict dominant transitions along the extinction pathway.
} 
\label{fig:ch2-gen}
\end{figure*}

\subsubsection{General case ($\mu>0$)}

By considering the other possibilities for dominant microscopic rates, we can map out the full behaviour of this model.  For $\mu<0$ we find that the $4\GG$ state is never a local minimum of $f$ so we do not consider this case in the following. For the same reason we do not consider the parameter regime $2J_R < -\mu$. 
The behaviour for other parameters is summarised in Fig.~\ref{fig:ch2-gen}, including the dominant microscopic transition pathways and the free energy profiles.

It turns out that $f(2\GG)\leq f(0\GG)$ throughout the regimes considered, so that $f^*=f(0\GG)$ and there is still no meaningful Allee effect [it remains type 0/I according to \eqref{eq:toy_allee_classification}].  Also,
 for any given $(\mu,J_R)$, the free energies $f(2\GG)$ and $f(0\GG)$ are piecewise linear functions of $J_H$.
Fig.~\ref{fig:ch2-gen}A shows the behaviour of $f(\text{0G})$, whose dependence on $J_H$ and $J_R$ is fully determined by two scaling variables $J_H/\mu$ and $J_R/\mu$.  The piecewise-linear dependence on $J_H$ is apparent, one also sees that $\mu\leq f^*\leq 2\mu$.  
There are four possibilities for the dominant pathways to extinction, which are shown in Fig.~\ref{fig:ch2-gen}B, in a representation similar to Fig.~\ref{fig:ch2-illust}.  
The sign of $J_R$ determines which is the dominant transition from 4G to 2G, while both $J_R$ and $J_H$ affect the outgoing transition.

Recalling that $f^*$ determines the MTE via \eqref{equ:tau-f*}, we see from Fig.~\ref{fig:ch2-gen}A that inhomogeneous mixing can only slow down extinction, as already discussed for the illustrative case of Fig.~\ref{fig:ch2-illust}.  The reason is the same: $J_H\neq 0$ tends to suppress the probability of certain microstates within 2G, which suppresses extinction pathways passing through those states.
This result -- that inhomogeneity slow down extinction -- is opposite to the result of the logistic (field-theoretic) model in Sec.~\ref{sec:extinctionLM}.  However, this should not be surprising since the mechanisms for these changes are quite different, and it was already apparent from the (field-thoretic) Schl\"ogl model that there is no general reason that the MTE should be affected in any particular direction.  Instead, we conclude that the simplicity of the toy model and the logistic model means that each or them can only support one sign for this effect.  Similarly, the absence of any meaningful Allee effect in the toy model is likely due to its simple structure -- there is no reason to expect that Allee effects will not occur in general models of this type.

\subsection{Model variant (``square root rates'')}
\label{sec:sqrt}

\begin{figure}[tb]
\includegraphics[width=0.86\linewidth]{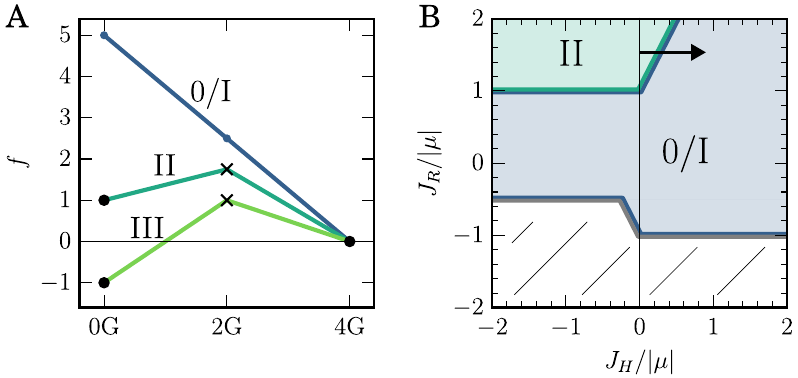}
\caption{ Analysis of Allee types for the model variant with $\chi_\beta(x) = e^{-\beta x/2}$ for $\mu > 0$.
\textbf{A}:  Effective free energies illustrating the Allee type classification \eqref{eq:toy_allee_classification}. Parameters: $\mu = [2.5,0.5,-0.5], J_R = [0,1,2], J_H = [0,0,0]$ for 0/I, II and III respectively.
\textbf{B}: Diagram showing how the Allee type depends on $J_H,J_R$, for $\mu>0$.  The white (hatched) regions shows parameters for which 4G is not a local minimum of $f$.  The black arrow indicates how a change in $J_H$ from homogeneous to inhomogeneous mixing can change the Allee type.   As discussed in the main text, the situation for $\mu<0$ is that type-II region becomes type-III, and the 0/I region becomes white (hatched) because $4\GG$ is no longer a local minimum of $f$.
}
\label{fig:ch2-sqrt}
\end{figure}

As a postscript to this analysis of the toy model, we show a slight change in its definition can produce a free energy barrier between 0G and 4G, which is a form of Allee effect.
Recall that detailed balance with respect to $\pi_R$ does not fully determine the transition rates in this model, and that other choices for the function $\chi_\beta$ are possible in (\ref{eq:mlmsdainwuqbdd},\ref{eq:mlmsdainwuqbdd2}).  An intuitive choice  might be a Metropolis-like rate $\chi_\beta={\rm min}[1,{\rm e}^{-\beta x}]$ but this yields the same low-temperature behaviour as found above for Glauber rates.  
Here we consider instead
$\chi_\beta(x)={\rm e}^{-\beta x/2}$
which is particularly convenient because it simplifies some formulae for the mesoscopic rates.  Note however that this choice for $\chi$ means that some microscopic rates can then diverge in the low-temperature limit, unlike to the original (Glauber-like) rates which are always less than unity.  This means in particular that the physically intuitive result (\ref{equ:tau-f*},\ref{equ:def-f*}) for the extinction time no longer holds.  

Still, it is instructive to compute the effective free energy within this model, using the method of Sec.~\ref{sec:projection2}. One finds
\begin{align}
\label{eq:cxmzdiwffbgbf}
\begin{split}
  f(0\text{G}) &= 2\mu\,, \\
  f(2\text{G}) &= \mu + J_R - f_{\text{inh}}(J_R,J_H)\,, \\
  f(4\text{G}) &= 0\,,
\end{split}
\end{align}
where  
\begin{align}
\label{eq:mdsjakndjwnqwdq}
 f_{\text{inh}}(J_R,J_H) =  \begin{cases}
   2J_H & 0 < 2 J_H \leq J_R \,,\\
   J_R & 0 < J_R\leq 2J_H \,,\\
   -J_R & 2J_H < J_R \leq 0 \,,\\
   -2J_H & J_R < 2J_H \leq 0 \,\\
   0 & \textrm{otherwise} \, .
 \end{cases}
\end{align}
This $f_{\rm inh}$ is non-negative and represents the difference in $f(2\GG)$ between the homogeneously mixed case ($J_H=0$) and the general free case [$f_{\rm inh} = f(2\GG) - f_0(2\GG)$].

In contrast to the toy model with Glauber rates, this effective free energy can support Allee types II and III, which are identified by barriers in $f$, that is $f(2\GG) > f(0\GG), f(4\GG)$.  Fig.~\ref{fig:ch2-sqrt}A illustrates examples of the types introduced in \eqref{eq:toy_allee_classification}.  Fig.~\ref{fig:ch2-sqrt}B shows how the Allee types depend on the model parameters for $\mu>0$.  We continue to restrict our analysis to cases where $4\GG$ is a local minimum of $f$. The white (hatched) region shows parameter regions for which 4G is not a local minimum.  The introduction of inhomogeneous mixing ($J_H\neq0$) can change the Allee type in this case (black arrow in Fig.~\ref{fig:ch2-sqrt}B). For $\mu<0$ the situation in Fig.~\ref{fig:ch2-sqrt}B is changed in that the type 0/I region becomes white (hatched) [because $f(4\GG)>f(2\GG)>f(0\GG)$] and the type II region changes to type III.

We end the discussion of this model variant by repeating that while these (effective) free energies are instructive, there is no simple analogue of \eqref{equ:tau-f*} in this model so the behaviour of the MTE cannot be deduced from \eqref{eq:cxmzdiwffbgbf}.  The MTE can be computed direct from the rates $r_{\rm meso}$ but we do not discuss that computation here.

\subsection{Discussion of toy model and connection to field-theoretic models}
\label{sec:discuss-toy}

The toy model discussed here is a very simple caricature of a system with separated time scales for hopping (migration) and chemical reactions (birth/death).  This section briefly summarises how it is useful for illustrating the fundamental assumptions and processes at work in the field-theoretical (AB-type) models  of Sec.~\ref{sec:modelab}.

The simple Markov chain shown in Fig.~\ref{fig:micro_statespace} shows how an irreversible non-equilibrium model can arise from hopping/reaction dynamics which separately respect detailed balance with respect to different energies.  This is analogous to the construction of Model AB.  By invoking a separation of time scales between hopping and reactions, this model may be projected onto a reaction co-ordinate where only the total populations of each species are relevant, and the dynamics of this co-ordinate follows a reversible (equilibrium-like) Markov chain.  This mirrors the projection of Model AB onto the global density, as accomplished in Sec.~\ref{sec:projection1}.  Nevertheless, both the field-theoretic model and the toy model show that the non-equilibrium structure of the underlying model has implications for the projected dynamics, for example that $F$ is not a function of state in Model AB, and that the dominant pathways between toy-model mesostates can be different for forward/backward transitions between the same mesostates of the toy model (recall Fig.~\ref{fig:ch2-illust}).

The low-temperature limit of the toy model is analogous to the small-$\epsilon$ limit of the AB-type model.  That is, extinction becomes a rare event and the MTE diverges exponentially fast, as shown by \eqref{equ:tau-f*} [analogous to \eqref{eq:djsabdwqbdqdwq}].  Also the transition to extinction concentrates on a specific path (instanton) which we were able to identify precisely within the toy model (Fig.~\ref{fig:ch2-gen}B).  The corresponding result for the AB-type model is that the density profiles along the instanton are always minimisers of $\mathcal{F}_B$, as shown in \eqref{equ:rhoB-min}.

We find that inhomogeneous mixing may affect the MTE in the toy model, although the mechanism of this effect differs from Model AB, and the MTE is only increased by inhomogeneity in the toy model, while changes of either sign can be found in Model AB (at least for the Schl\"ogl case).  Our interpretation of this last fact is that the simple toy model considered here cannot support the full range of phenomenology supported by non-equilibrium systems with slow reactions and fast migration, but more complex model variants might well allow richer behaviour.  (For example, the restriction to $\sum_m (g_m+2w_m)=4$ was an arbitrary choice, taken for simplicity.  Introducing more particles might well reveal new possible behaviours.)  Nevertheless, we find that the simple cases analysed here are already useful to illustrate the non-equilibrium aspects of these models, and how they impact on rare events like the MTE.

%% file: sections/bigtikz.tex
\makebox[\textwidth][c]{
\scalebox{1.2}{
\begin{tikzpicture}[-,>=stealth,shorten >=1pt,auto,scale=1,node distance=3cm,thick,main node/.style={circle,draw,font=\Large}]

\definecolor{colorA}{rgb}{0.369214, 0.788888, 0.382914}
\definecolor{colorB}{rgb}{0.99, 0.99, 0.99}



\node[rectangle,
    draw = gray, dashed,
    anchor = north west,
    minimum width = 12cm, 
    minimum height = 1.5cm] at (-2.5,0.8) {};
\node[rectangle,
    draw = gray, dashed,
    anchor = north west,
    minimum width = 12cm, 
    minimum height = 3.6cm] at (-2.5,-1.2) {};
\node[rectangle,
    draw = gray, dashed,
    anchor = north west,
    minimum width = 12cm, 
    minimum height = 1.5cm] at (-2.5,-6.2+0.5) {};

\node at (-3, 0) {$\equiv$};
\node at (-3,-3) {$\equiv$};    
\node at (-3,-6.5) {$\equiv$};

\node[rectangle,draw,minimum width=0.9cm,minimum height=0.9cm] (4A)  at (-4,0) {\textbf{4G}};
\node[rectangle,draw,minimum width=0.9cm,minimum height=0.9cm] (2A)  at (-4,-3) {\textbf{2G}};
\node[rectangle,draw,minimum width=0.9cm,minimum height=0.9cm] (0A)  at (-4,-6.5) {\textbf{0G}};


\node[rectangle,draw,minimum width=1.2cm,minimum height=0.6cm] (1) at (0,0) {};
\draw[thin] (0,-0.28) -- (0,0.32) ;
\node[rectangle,draw,minimum width=1.2cm,minimum height=0.6cm] (2) at (3.5,0) {};
\draw[thin] (3.5,-0.3) -- (3.5,0.32) ;
\node[rectangle,draw,minimum width=1.2cm,minimum height=0.6cm] (3) at (7,0) {};
\draw[thin] (7,-0.3) -- (7,0.32) ;

\node[rectangle,draw,,minimum width=1.2cm,minimum height=0.6cm] (6) at (0,-4) {};
\draw[thin] (0,-4-0.3) -- (0,-4+0.32) ;
\node[rectangle,draw,,minimum width=1.2cm,minimum height=0.6cm] (7) at (3.5,-2) {};
\draw[thin] (3.5,-2-0.3) -- (3.5,-2+0.32) ;
\node[rectangle,draw,,minimum width=1.2cm,minimum height=0.6cm] (8) at (7,-4) {};
\draw[thin] (7,-4-0.3) -- (7,-4+0.32) ;

\node[rectangle,draw,,minimum width=1.2cm,minimum height=0.6cm] (12) at (0,-6.5) {};
\draw[thin] (0,-6.5-0.3) -- (0,-6.5+0.32) ;
\node[rectangle,draw,,minimum width=1.2cm,minimum height=0.6cm] (13) at (7,-6.5) {};
\draw[thin] (7,-6.5-0.3) -- (7,-6.5+0.32) ;

\node[circle,fill=colorA,minimum size=5pt, inner sep=0pt] (c) at (-0.425,-0.12){};
\node[circle,fill=colorA,minimum size=5pt, inner sep=0pt] (c) at (-0.2,-0.12){};
\node[circle,fill=colorA,minimum size=5pt, inner sep=0pt] (c) at (-0.425,0.12){};
\node[circle,fill=colorA,minimum size=5pt, inner sep=0pt] (c) at (-0.2,0.12){};

\node[circle,fill=colorA,minimum size=5pt, inner sep=0pt] (c) at (3.5-0.425,-0.12){};
\node[circle,fill=colorA,minimum size=5pt, inner sep=0pt] (c) at (3.5-0.2,-0.12){};
\node[circle,fill=colorA,minimum size=5pt, inner sep=0pt] (c) at (3.5-0.3125,0.12){};
\node[circle,fill=colorA,minimum size=5pt, inner sep=0pt] (c) at (3.5+0.3,-0.12){};

\node[circle,fill=colorA,minimum size=5pt, inner sep=0pt] (c) at (7-0.425,-0.12){};
\node[circle,fill=colorA,minimum size=5pt, inner sep=0pt] (c) at (7-0.2,-0.12){};
\node[circle,fill=colorA,minimum size=5pt, inner sep=0pt] (c) at (7+0.425,-0.12){};
\node[circle,fill=colorA,minimum size=5pt, inner sep=0pt] (c) at (7+0.2,-0.12){};

\node[circle,fill=colorA,minimum size=5pt, inner sep=0pt] (c) at (-0.425,-4-0.12){};
\node[circle,fill=colorA,minimum size=5pt, inner sep=0pt] (c) at (-0.2,-4-0.12){};
\node[circle,fill=colorB,minimum size=5pt, inner sep=0pt, line width=0.3pt, draw=black] (c) at (-0.3125,-4+0.12){};

\node[circle,fill=colorA,minimum size=5pt, inner sep=0pt] (c) at (3.5-0.425,-2-0.12){};
\node[circle,fill=colorB,minimum size=5pt, inner sep=0pt, line width=0.3pt, draw=black] (c) at (3.5-0.2,-2-0.12){};
\node[circle,fill=colorA,minimum size=5pt, inner sep=0pt] (c) at (3.5+0.3125,-2-0.12){};

\node[circle,fill=colorB,minimum size=5pt, inner sep=0pt, line width=0.3pt, draw=black] (c) at (7-0.3125,-4-0.12){};
\node[circle,fill=colorA,minimum size=5pt, inner sep=0pt] (c) at (7+0.425,-4-0.12){};
\node[circle,fill=colorA,minimum size=5pt, inner sep=0pt] (c) at (7+0.2,-4-0.12){};

\node[circle,fill=colorB,minimum size=5pt, inner sep=0pt, line width=0.3pt, draw=black] (c) at (-0.425,-6.5-0.12){};
\node[circle,fill=colorB,minimum size=5pt, inner sep=0pt, line width=0.3pt, draw=black] (c) at (-0.2,-6.5-0.12){};

\node[circle,fill=colorB,minimum size=5pt, inner sep=0pt, line width=0.3pt, draw=black] (c) at (7-0.3,-6.5-0.12){};
\node[circle,fill=colorB,minimum size=5pt, inner sep=0pt, line width=0.3pt, draw=black] (c) at (7+0.3,-6.5-0.12){};

\draw[-{Stealth[left]}] ([yshift=0pt,xshift=1pt]4A.south) -- ([yshift=-1pt,xshift=1pt]2A.north) node[midway,left] {}; 
\draw[-{Stealth[left]}] ([yshift=0pt,xshift=-1pt]2A.north) -- ([yshift=1pt,xshift=-1pt]4A.south) node[midway,right] {};
\draw[-{Stealth[left]}] ([yshift=0pt,xshift=1pt]2A.south) -- ([yshift=-1pt,xshift=1pt]0A.north) node[midway,left] {}; 
\draw[-{Stealth[left]}] ([yshift=0pt,xshift=-1pt]0A.north) -- ([yshift=1pt,xshift=-1pt]2A.south) node[midway,right] {};

\draw[-{Stealth[left]}] ([yshift=1pt]1.east) -- ([yshift=1pt,xshift=2pt]2.west) node[midway,above] {$4$}; 
\draw[-{Stealth[left]}] ([yshift=-1pt]2.west) -- ([yshift=-1pt,xshift=-2pt]1.east) node[pos=0.46,below] {1}; 

\draw[-{Stealth[left]}] ([yshift=1pt]2.east) -- ([yshift=1pt,xshift=2pt]3.west) node[midway,above] {$3$}; 
\draw[-{Stealth[left]}] ([yshift=-1pt]3.west) -- ([yshift=-1pt,xshift=-2pt]2.east) node[pos=0.46,below] {2};

\draw[-{Stealth[left]},line width=0.5mm] ([yshift=1.5pt,xshift=0pt]6.north east) -- ([yshift=1.5+1pt,xshift=1pt]7.south west) node[pos=0.75,above left,rotate=30] {$2\chi_\beta(-J_H)$}; 
\draw[-{Stealth[left]}] ([yshift=-1.5pt,xshift=0pt]7.south west) -- ([yshift=-1.5-1pt,xshift=-1pt]6.north east) node[pos=0.7,below right,rotate=30] {$\chi_\beta( J_H)$}; 

\draw[-{Stealth[left]},line width=0.5mm] ([yshift=1.5pt,xshift=0pt]7.south east) -- ([yshift=1.5-1pt,xshift=1pt]8.north west) node[pos=0.2,above right,rotate=330] {$\chi_\beta(- J_H)$}; 
\draw[-{Stealth[left]}] ([yshift=-1.5pt,xshift=0pt]8.north west) -- ([yshift=-1.5+1pt,xshift=-1pt]7.south east) node[pos=0.25,below left,rotate=330] {$2\chi_\beta( J_H)$}; 

\draw[-{Stealth[left]},line width=0.7mm] ([yshift=1.5pt,xshift=0pt]6.east) -- ([yshift=1.5pt,xshift=2pt]8.west) node[pos=0.5,above] {$\chi_\beta(-2J_H)$}; 
\draw[-{Stealth[left]}] ([yshift=-1.5pt,xshift=0pt]8.west) -- ([yshift=-1.5pt,xshift=-2pt]6.east) node[pos=1-0.5,below] {$\chi_\beta(2J_H)$};

\draw[-{Stealth[left]}] ([yshift=1pt,xshift=0pt]12.east) -- ([yshift=1pt,xshift=2pt]13.west) node[midway,above] {2}; 
\draw[-{Stealth[left]}] ([yshift=-1pt,xshift=0pt]13.west) -- ([yshift=-1pt,xshift=-2pt]12.east) node[pos=0.485,below] {1}; 

\draw[-{Stealth[left]}] ([yshift=0pt,xshift=1.5pt]1.south) -- ([yshift=-1pt,xshift=1.5pt]6.north) node[midway,right] {$12 k \chi_\beta(2J_R + \mu)$}; 
\draw[-{Stealth[left]},line width=0.7mm] ([yshift=0pt,xshift=-1.5pt]6.north) -- ([yshift=1pt,xshift=-1.5pt]1.south) node[midway,left] {$k \chi_\beta(-2J_R - \mu)$};  

\draw[-{Stealth[left]}] ([yshift=0pt,xshift=1.5pt]2.south) -- ([yshift=-1pt,xshift=1.5pt]7.north) node[pos=0.45,right] {$6k \chi_\beta(J_R + \mu)$}; 
\draw[-{Stealth[left]},line width=0.5mm] ([yshift=0pt,xshift=-1.5pt]7.north) -- ([yshift=1pt,xshift=-1.5pt]2.south) node[midway,left] {$k \chi_\beta(-J_R - \mu)$};  

\draw[-{Stealth[left]}] ([yshift=0pt,xshift=1pt]3.south) -- ([yshift=-1pt,xshift=1pt]8.north) node[midway,right] {$2k \chi_\beta(\mu)$}; 
\draw[-{Stealth[left]}] ([yshift=0pt,xshift=-1pt]8.north) -- ([yshift=1pt,xshift=-1pt]3.south) node[midway, left] {$k \chi_\beta(-\mu)$};  

\draw[-{Stealth[left]},line width=0.7mm] ([yshift=0pt,xshift=1.5pt]6.south) -- ([yshift=-1pt,xshift=1.5pt]12.north) node[midway,right] {$2k \chi(-2J_R+\mu)$};
\draw[-{Stealth[left]}] ([yshift=0pt,xshift=-1.5pt]12.north) -- ([yshift=1pt,xshift=-1.5pt]6.south) node[midway,left] {$2k \chi_\beta(2J_R-\mu)$};

\draw[-{Stealth[left]}] ([yshift=0pt,xshift=1pt]8.south) -- ([yshift=-1pt,xshift=1pt]13.north) node[midway, right] {$2k \chi_\beta(\mu)$};
\draw[-{Stealth[left]}] ([yshift=0pt,xshift=-1pt]13.north) -- ([yshift=1pt,xshift=-1pt]8.south) node[midway, left] {$k \chi_\beta(-\mu)$};

\draw[<->,line width=0.5pt] (1,-7.5) -- (6,-7.5) node[midway, below] {fast};

\draw[->,line width=0.5pt] (10,0) -- (10,-6.5) node[midway, below, rotate=90] {slow};

\end{tikzpicture}
}
}

%% file: sections/conclusion.tex
\section{Conclusion}
\label{sec:conclusion}

This section briefly summarises our main conclusions, together with an outlook.  (We refer back to Secs.~\ref{sec:discuss-AB} and~\ref{sec:discuss-toy} for additional remarks on the AB-type and toy models respectively.)
For both field-theoretic and toy models, we have shown how MTEs can be computed analytically within appropriate limits, which require both weak noise (so that extinction is a rare event) and a separation of time scales between fast diffusion and slow reactions (so that the dynamics can be projected onto a single co-ordinate).  
This enables construction of an effective free energy for this co-ordinate, from which the weak-noise scaling of the MTE can be deduced, even though the system is far from equilibrium.
Within this framework, Allee effects play a natural role: we have classified these into various types according to their effect on the MTE and their relation with the (effective) free energies $F,f$ of the two models: this extends the most common classification~\cite{stephens1999,dennis2002,gastner2011,meerson2011} which only depends on on the per-capita population growth rates [for example $R(\varphi)/\varphi$].

For the AB-type models, the possibility of phase separation enters the effective free energy only through its binodal densities, and the values of the chemical potential and mobility at those points.  This simplification is due to the time scale separation.  In general, phase separation may increase or reduce the MTE although we identified conditions on the concavity/convexity of net reaction rate and mobility under which the MTE can only be reduced.  These conditions are satisfied in the logistic model but not in the Schl\"ogl model.  The Allee type can also be modified by phase separation in the Schl\"ogl model but this is not possible in the logistic model, which we again attribute to certain special features of that case (specifically, that the per-capita growth rate is monotonic).  The modification of Allee type illustrates how particle-hopping (diffusive) dynamics can have qualitative effects on the dynamics of the total population, despite the fact that diffusion does not directly create or destroy particles.

For the toy model, it turns out the MTE can only be increased by inhomogeneous mixing.  Also, while we are able to define meaningful analogues of Allee effects, these do not appear in the model as originally defined (with Glauber rates).  We attribute both of these aspects to the simplicity of the model, but we emphasise that it still illustrates the interplay between non-equilibrium systems (non-reversible Markov chains), weak noise, time scale separation, and rare events.

Our results complement existing results for Model AB~\cite{li2020,li2021} and extend those works to characterisation of rare events.  The possibility of extinction events in such models further highlights their rich phenomenology, although we have emphasized that deriving them from underlying particle systems requires extra care if one wants to accurately capture the probability of rare events such as extinction~\cite{doering2005}.  A central assumption enabling our analysis is the complete separation of time scales between fast diffusive dynamics and slow birth/death. 

 It would be very interesting to understand how our conclusions are modified by slow processes within the diffusive sector, which might still be theoretically tractable (for example, nucleation occurring by a Model-B-like mechanism might be slow even if hopping is fast).  More generally, one may imagine other systems with spatially-extended population dynamics  where rare events can be analysed by projection onto one or two reaction co-ordinates, in order to render them tractable.  Such co-ordinates might decouple from other dynamical modes via a timescale separation (as in this study) or via symmetry considerations, as discussed for a model with conserved density in~\cite{cates2023}.

%% file: sections/appendix.tex
\section{Projection onto reaction coordinate in discrete model AB}
\label{sec:discrete_ab}

In this appendix, we derive analogous results to Sec.~\ref{sec:modelab} for a discrete particle-based model, instead of the AB-type model that involves continuous densities \eqref{eq:modelab}. 
The essential differences come from the use of Gaussian and Poissonian noises in the two cases~\cite{doering2005}.
The main text considers systems driven by Gaussian noise, and while the discrete models considered here have  Poissonian noise. 
In this context, it is important that the continuous models with Gaussian noise are derived from discrete systems using central limit theorems (CLTs), which is sufficient to ensure that the typical behaviour matches between continuous and discrete, but it does not ensure that large-deviation properties will be the same.  This means in particular that the free energy of the Gaussianized (continuous) model will not generally match that of the discrete one.  Still, the qualitative behaviour of the different models is very similar, for the cases considered here.

Diagrammatically we have:
\begin{center}
\begin{tikzpicture}

\node[rectangle,draw,minimum width=1.2cm,minimum height=0.6cm] (1) at (0,0) {discrete Model $\ve{n}$, Eq. \eqref{eq:modelab_discrete}};
\node[rectangle,draw,minimum width=1.2cm,minimum height=0.6cm] (2) at (9,0) {continuous Model $\rho(x)$, Eq. \eqref{eq:modelab}};
\node[rectangle,draw,minimum width=1.2cm,minimum height=0.6cm] (3) at (0,-2) {free energy $G(n)$};
\node[rectangle,draw,minimum width=1.2cm,minimum height=0.6cm] (4) at (9,-2) {free energy $F(\rho)$};
\node[rectangle,draw,minimum width=1.2cm,minimum height=0.6cm] (5) at (5.5,-2) {free energy $\tilde{F}(\rho)$};
\node[] (6) at (7.28,-2) { {\Large $\neq$ }};

\draw[->,>=stealth,thick] (1.east) -- (2.west) node[midway,above] {via \eqref{eq:skellamlimit}}; 
\draw[->,>=stealth,thick] (3.east) -- (5.west) node[midway,above] {via \eqref{eq:F_R} }; 
\draw[->,>=stealth,thick] (1.south) -- (3.north) node[midway,right] {via \eqref{eq:discrete_G}}; 
\draw[->,>=stealth,thick] (2.south) -- (4.north) node[midway,right] {via \eqref{eq:F_A}}; 

\end{tikzpicture}
\end{center}
where \eqref{eq:skellamlimit} is the relevant central limit theorem, while (\ref{eq:F_R},\ref{eq:F_A}) require estimation of probabilities of rare (large-deviation events).
In Sec.~\ref{sec:modelab} we discussed the right side of the diagram. In this appendix, we complement Sec.~\ref{sec:modelab} by discussing the left side, including derivation of $\tilde{F}$.

\subsection{Model}

We consider particles positioned on a lattice with $M$ lattice sites labelled $1,2,\ldots,M$. A lattice site $i$ has $N_i(t)$ particles and the system has state $\ve{N}(t) = (N_1(t),N_2(t),\ldots, N_M(t))$ at time $t$. Particles can undergo chemical reactions by which they can increase or decrease the particle numbers, and hops between lattice sites. We assume reactions to be local on a lattice site. 

For a site with occupancy $n$, we assume the chemical reactions form a birth-and-death process with birth rate $w^\mathrm{b}(n)$ and death rate $w^\mathrm{d}(n)$.   Throughout this appendix, upper case symbols $N,\Delta,\Delta N$ indicate random variables, in particular $N_i(t)$ is the (random) number of particles on site $i$.  We use lower case $n_i$ to indicate possible values of the random variable.
The (random) change due to reactions of the occupancy of site $i$ in a short time $\Delta t$ is $\Delta N_i^R(t)$, which follows a Skellam distribution, the difference of two Poisson processes. That is
\begin{align}
\label{eq:N_R}
  \Delta N^\mathrm{R}_i(t) &= \Delta^\mathrm{R,b}_{i}(t) -  \Delta^\mathrm{R,d}_{i}(t) \; ,
\end{align}
where the independent increments for birth and death are $\Delta^\mathrm{R,b}_{i}$ and $\Delta^\mathrm{R,d}_{i}$, which are Poisson distributed
\begin{align}
\begin{split}
  \Delta^{\mathrm{R},\alpha}_{i}(t) |_{\ve{N}(t) = \ve{n}} &\sim \mathrm{Poi} ( \Delta t \, w^\alpha (n_i) ) \; ,
\end{split}
\end{align}
where $\alpha\in\{{\rm b},{\rm d}\}$, also $\bm{n}=(n_1,n_2,\dots,n_M)$ and 
the notation $X|_{A} \sim \mathrm{Poi}(\lambda)$ indicates that the random variable $X$ has a Poissonian conditional distribution $P(X=k|A) = \frac{\lambda^k}{k!} e^{-\lambda}$, where $A$ is the event on which we are conditioning.

The hopping dynamics are described similarly: 
write $J_{i-1\to i}$ for the (net) current from site $i-1$ to $i$ in a short time $\Delta t$.
It is the difference between the right and left-moving fluxes across the bond, leading to a similar Skellam distribution
\begin{equation}
\begin{split}
  J_{i-1\to i}(t) & = \Delta^\mathrm{D}_{i-1 \to i}(t) - \Delta^\mathrm{D}_{i \to i-1}(t) \,,
  \\
    \Delta^\mathrm{D}_{i \to i \pm 1}(t) |_{\ve{N}(t) = \ve{n}} &\sim \mathrm{Poi} ( \Delta t \, k_{i \to i \pm 1} (\ve{n}) ) \,,
\end{split}
\end{equation}
where $k_{i\to i\pm 1}(\bm{n})$ is a (suitably normalised) hop rate from site $i$ to site $i\pm1$. 
These hopping rates respect detailed balance with respect to a suitable grand-canonical distribution, for example one may take
\begin{equation}
 k_{i \to i \pm 1}(\ve{n})  = k_0 n_i
\exp\left[-\frac{\beta}{2} \left(E(\ve{n} + \ve{e}^{i \pm 1} - \ve{e}^i) - E(\ve{n})\right) \right]
\end{equation}
where $k_0$ is a bare hopping rate, $\ve{e}^i$ is a vector of size $M$ with $(\ve{e}^i)_j = \delta_{ij}$, and $E=E(\ve{n})$ is an energy function that describes particle interactions, see below for further discussion.  
We take periodic boundaries, for simplicity.

Together, the (stochastic) change in occupation $\Delta N_i(t)$ is
\begin{align}
\label{eq:modelab_discrete}
\begin{split}
  \Delta N_i(t) &= \Delta N^\mathrm{R}_i(t) - (\mathrm{div}\, J(t))_i
\end{split}
\end{align}
where $\mathrm{div}$ is the discrete divergence operator, that is $(\mathrm{div}\, J)_i = J_{i\to i+1} - J_{i-1\to i}$.
Eq.~\eqref{eq:modelab_discrete} is the discrete analogue of Eq.~\eqref{eq:modelab}.

As in the main text, we assume that the interaction energy $E$ for the hopping dynamics contains attractions that will drive phase separation.  The total number of particles is $N_{\rm tot}(t) = \sum_i N_i(t)$ and there exist binodals such that for $n_- < (N_{\rm tot}(t)/M) < n_+$ the hopping dynamics drives phase separation into regions where the local occupancies are (close to) $n_-$ and $n_+$.  As in the main text, this diffusive dynamics is assumed to be fast compared to the reactions, in which case the details of the energy $E$ do not affect the results, except through the binodal densities.

\subsection{Single-site free energy}
\label{app:free}

The next step is to compute the behaviour of the reaction dynamics on a single site $i$.
The (scaled) mean and variance of the $\Delta N_i(t)$ (conditioned on the current state ${N}_i={n}_i$) are
\begin{align}
\label{eq:eq:discrete_reaction_avg}
\begin{split}
  r^\mathrm{R}(n_i) &:= \frac1{\Delta t} \langle \Delta N^\mathrm{R}_i | N_i = n_i \rangle  =   w^\mathrm{b}(n_i) - w^\mathrm{d}(n_i)  \\
  2 m^\mathrm{R}(n_i) &:= \frac1{\Delta t} \mathrm{Var}( \Delta N^\mathrm{R}_i | N_i = n_i ) =  w^\mathrm{b}(n_i) + w^\mathrm{d}(n_i) 
\end{split}
\end{align}
These results inspired the form of the reaction rate and mobility of the AB-type model in Eqs.~(\ref{eq:MA-muA},\ref{eq:RA}). 
If $w^\mathrm{b}(k)$ and $w^{\rm d}(k+1)$ are positive for all $k\geq0$, the system relaxes to a steady state  
\begin{equation}
\pi^\mathrm{R}(n) \propto  \prod_{k=0}^{n-1}  \frac{w^\mathrm{R,b}(k)}{w^\mathrm{R,d}(k+1)} \,,
\label{equ:piR}
\end{equation}
where the proportionality constant is set by normalisation.
The carrying capacity is defined as $n^*=\operatorname{argmax}_n \pi(n)$.  Then we define a (discrete) free energy as
\begin{align}
\label{eq:discrete_G}
  G^\mathrm{R}(n) = -\log\frac{\pi^\mathrm{R}(n)}{\pi^{\mathrm{R}}(n^*)}  
\end{align}
which vanishes at the carrying capacity.

To illustrate this theory, consider
the discrete logistic model as discussed in Sec.~\ref{sec:CBDK} which is defined by birth and death rates
\begin{equation}
  w^\mathrm{b}(n) = n, \quad w^\mathrm{d}(n) = n(n-1)/n_s\,,
\end{equation}
where  $n_s>0$ is a parameter.  Note that the number of particles is constrained in this model as $n\geq 1$.
The particle number has a Poisson distribution conditioned on $n\geq 1$, so the carrying capacity is $n^* = \lfloor{n_s}\rfloor$ and
\begin{equation}
  G^\mathrm{R}(n) =  (n-n^*)\log n_s - \log \frac{n!}{n^*!} \,.
\end{equation}
As discussed in Sec.~\ref{sec:CBDK}, it is convenient when studying extinction to supplement this model with an extra absorbing state $N=0$ such that a system with $N=1$ transitions to $N=0$ with some small rate $\epsilon_X$, after which it remains extinct for ever.  If $\epsilon_X$ is small enough then the system will relax to the metastable distribution $\pi$ before becoming extinct, and we identify the overall extinction rate as $\epsilon_X \pi(1)$.

\subsection{From discrete to continuous density: large-$K$ limit}
\label{app:largeK}

We now follow~\cite{meerson2011,assaf2010,doering2005} and consider a limit where the typical occupancy of each site is very large.  Specifically, we introduce a large parameter $K\gg 1$ such that the discrete occupancies $N_i \sim K$ and we define densities $\rho_i = N_i/K$ which we can approximate for large $K$ as continuous random variables.  To achieve $N_i\sim K$ we impose that birth and death rates scale with $K$ as 
\begin{equation}
\label{equ:wa-K}
w^{\alpha}(n) = K \lambda_\alpha(n/K)
\end{equation}
where $\alpha\in\{\mathrm{b},\mathrm{d}\}$ indicates either birth or death, and $\lambda_{\rm b},\lambda_{\rm d}$ are smooth functions. In the example of the logistic model, these functions are
\begin{equation}
  \lambda_\mathrm{b}(\rho) = \rho, \quad \lambda_\mathrm{d}(\rho) = \rho^2/\rho_s \,,
\end{equation}
where the new parameter $\rho_s$ is related to the original carrying capacity via $n_s = K\rho_s$.

As anticipated above, taking $K\to\infty$ in this way enables a central limit theorem in which the continuous dynamics of $\rho_i$ may be approximated by a stochastic differential equation.   Recalling that Poisson distributions for large argument may be approximated by appropriate Gaussians, we note for large $K$ and fixed $a=O(1)$ that
\begin{equation}
\label{eq:skellamlimit}
(1/K) \mathrm{Poi}(Ka) \to \mathcal{N} (a, a/K ) 
\end{equation}
Using this together with \eqref{eq:eq:discrete_reaction_avg}, 
we find that as $K \to \infty$ 
\begin{equation}
\label{eq:ndioqnmdsakdnsa}
  \frac1K \Delta N^\mathrm{R}_i(t)  \to  -M^\mathrm{R}(\rho_i) \mu^R(\rho_i) \Delta t  + \sqrt{2 M^\mathrm{R}(\rho_i)/K} \, \Delta W^\mathrm{R}(t)
\end{equation}
where 
$M^\mathrm{R}(\rho) = m^\mathrm{R}(K \rho) / K$, $\mu^\mathrm{R}(\rho) = r^\mathrm{R}(K \rho) / m^\mathrm{R}(K\rho)$,
and $\Delta W^\mathrm{R}(t) \sim {\cal N}(0,\Delta t)$ is a suitable increment of a Wiener process. 

Identifying $\Delta N^\mathrm{R}_i(t)/K$ as the change in $\rho_i$ due to reactions and $\epsilon=1/K$, this equation reduces to a spatially-discretised form of the model-A parts of \eqref{eq:modelab}, that is
\begin{equation}
\partial_t \rho_i = -M^\mathrm{R}(\rho_i) \mu^R(\rho_i)  + \sqrt{2 \epsilon M^\mathrm{R}(\rho_i)} \, \Lambda^R_i(t)
\label{equ:rho-sde}
\end{equation}
With appropriate choices for $E$ the currents $J_{i\to i-1}$ for the particle-hopping dynamics can also be mapped to a discretised version of the model-B parts of \eqref{eq:modelab} (perhaps with density-dependent noise terms), but we do not consider that computation here since the model-B dynamics only enter the dynamics through the binodals $n_\pm$, as described above.

We emphasise again that \eqref{equ:rho-sde} was derived by applying a central limit theorem to the discrete particle model, leading to a Gaussian noise.  We therefore refer to it as a ``Gaussianised'' model.

\subsection{Free energies for discrete and Gaussianised models}
\label{app:FR}

In the large-$K$ limit, the behaviour of the free energy $G$ of the discrete model can be characterised  from (\ref{equ:piR},\ref{eq:discrete_G},\ref{equ:wa-K}) as
\begin{align}
\label{eq:F_R}
  \tilde{F}_\mathrm{R}(\rho) := \lim_{K\to \infty} \frac1K G^\mathrm{R}(\rho K) = -\int_{\rho^*}^\rho \dif\rho'\, \log\frac{\lambda_\mathrm{b}(\rho')}{\lambda_\mathrm{d}(\rho')} \,.
\end{align}
where $\rho^*$ is the zero of the free energy (analogous to $n^*$), which we take to be its global minimum, where $\lambda_{\rm b}=\lambda_{\rm d}$.  
This means that the asymptotic distribution $\pi^{\rm R}$ behaves asymptotically as $\pi^{\rm R}(\rho K) \simeq {\rm e}^{-K\tilde{F}_R(\rho)}$ for large $K$.

However, applying the free-energy computation of Sec \ref{sec:ABtype} to the Gaussianised model \eqref{equ:rho-sde} one obtains instead that the distribution of $\rho$ behaves  as $P_{\rm A}(\rho)\simeq {\rm e}^{-K{F}_A(\rho)}$ with
\begin{equation}
\label{eq:dnqdqiowdnqwd2}
  F_\mathrm{A}(\rho) = -2\int_{\rho^*}^\rho\dif\rho'\, \frac{\lambda_\mathrm{b}(\rho') - \lambda_\mathrm{d}(\rho')}{\lambda_\mathrm{b}(\rho') + \lambda_\mathrm{d}(\rho')} \,,
\end{equation}
analogous to Eqs. \eqref{eq:MA-muA} and \eqref{eq:F_A}.
Clearly $F_\mathrm{A}\neq \tilde{F}_\mathrm{R}$ in general, so the full distribution of $\rho$ in the Gaussianised model does not match  its discrete counterpart, although it can be verified that they have the same minima [characterised by $ \lambda_\mathrm{b}(\rho^*)= \lambda_\mathrm{d}(\rho^*)$] and that they have the same curvatures there $F_{\rm A}''(\rho^*) = F_{\rm R}''(\rho^*)$.  To understand the differences between them, we recall that Gaussianising the noise using the central limit theorem \eqref{eq:skellamlimit} does not capture the rare events that eventually determine the tails of $\pi^{\rm R}$.  Hence $F_{\rm A}$ does not accurately describe large density fluctuations of the discrete particle model.  On the other hand, the qualitative features of the discrete model are largely captured by its Gaussianised version.

As an instructive example, 
the logistic model free energy according to \eqref{eq:F_R} is
\begin{equation}
  \tilde{F}^\mathrm{R}(\rho) = -\rho + \rho_s + \rho\log\frac{\rho}{\rho_s}
\end{equation}
which is to be compared with the free energy of the Gaussianised model: $F_\mathrm{A}(\rho) = 2\left(\rho - \rho_s - 2\rho_s \log\left(\frac{\rho+\rho_s}{2\rho_s}\right)\right)$ as obtained from \eqref{eq:dnqdqiowdnqwd2}, see also Eq.  \eqref{eq:dsandqwhudehfgiobjfv} in the main text.  The Schlögl free energy according to \eqref{eq:F_R} is given by
\begin{equation}
  F^\mathrm{R}(\rho) =  \rho_s-\rho  - \rho  \log\frac{\rho (\rho_s+\rho_u)}{\rho^2 + \rho_s\rho_u} + 2\sqrt{\rho_s\rho_u}\arctan\left(\sqrt{\frac{\rho_u}{\rho_s}} \frac{\rho - \rho_s}{\rho + \rho_u} \right) 
\end{equation}
which is to be compared with $F_A$ in Eq. \eqref{eq:Njndbashdwyd}, which was again derived using \eqref{eq:dnqdqiowdnqwd2}.

\subsection{Projection}

We now project the dynamical equation \eqref{eq:modelab_discrete} onto the total number of particles in the system
\begin{equation}
  N_{\rm tot} = \sum_i N_i
\end{equation}
which we choose as reaction coordinate. 
Apart from minor technical adjustments, the following derivation mirrors the one in Sec.~\ref{sec:projection1}, although we focus on 
the total number of particles instead of the mean density. 

In a phase-separated state, the two phases sit at binodals $n_i \approx n_\pm$. The dense (dilute) phase comprises of $M_+$ ($M_-$) lattice sites. 
We assume a large system such that we can ignore interface effects and $M_++M_-=M$, so that
\begin{equation}
\label{eq:mssajnd3dfoiwfndabasd}
  M_+ = \frac{N_{\rm tot} - M n_-}{n_+ - n_-}, \quad M_- = \frac{M n_+ - N_{\rm tot}}{n_+ - n_-} \,.  
\end{equation}
To derive a dynamical equation for the total particle number $N_{\rm tot}$, we start by summing Eq. \eqref{eq:modelab_discrete} over all lattice sites. 
The increment $\Delta N_{\rm tot}(t) := N_{\rm tot}(t+\Delta t) - N_{\rm tot}(t)$  obeys
\begin{equation}
\Delta N_{\rm tot}(t)
= \sum_i \Delta N_i(t) = \sum_i \Delta N_i^R(t) \,.
\end{equation}
where the second equality used that $\sum_i (\operatorname{div} J)_i=0$.
Using \eqref{eq:N_R} one obtains
\begin{equation}
\label{eq:njsdabdaudbwu3efrghtgf}
\Delta N_{\rm tot} = \sum_i  (\Delta^b_{i} -  \Delta^d_{i}) = \Delta^b -  \Delta^d
\end{equation}
in which $\Delta^\alpha := \sum_i \Delta_{i}^\alpha$ are distributed as
\begin{equation}
\Delta^\alpha(t) |_{N_{\rm tot}(t)=n_{\rm tot}}  \sim \mathrm{Poi} ( \Delta t \,  w_\mathrm{PS}^\alpha ({n}_{\rm tot}) )
\end{equation}
where 
$w_\mathrm{PS}^\alpha(n_{\rm tot}) = \sum_i w^\alpha(n_i) $ only depends on the total occupancy $n_{\rm tot}=\sum_i n_i$, as we now explain.
We assume that the hopping dynamics is fast enough that the system is either homogeneous or phase-separated.
In a homogeneous state then $n_i\approx n_{\rm tot}/M$ for all $i$ so we may approximate 
\begin{equation} 
w_\mathrm{PS}^\alpha(n_{\rm tot}) = M w^\alpha(n_{\rm tot}/M) \,.
\label{equ:wPS-hom}
\end{equation}
On the other hand, in a phase separated state we can approximate sums over lattice sites using Eq. \eqref{eq:mssajnd3dfoiwfndabasd} to obtain 
\begin{equation}
 w_\mathrm{PS}^\alpha(n_{\rm tot})
  = \frac{n_{\rm tot} - M n_-}{n_+ - n_-} w^\alpha(n_+) + \frac{M n_+ - n_{\rm tot}}{n_+ - n_-} w^\alpha(n_-)  \,.
\label{equ:wPS-sep}
\end{equation}

The stochastic process for $N_{\rm tot}(t)$ in \eqref{eq:njsdabdaudbwu3efrghtgf} becomes a birth-and-death process which can be analysed similarly to Sec.~\ref{app:free}.
Specifically, the steady-state distribution is 
\begin{equation}
\pi_{\rm PS}(n_{\rm tot}) \propto \prod_{i=0}^{n_{\rm tot}-1} \log\frac{ w_\mathrm{PS}^\mathrm{b}(k)}{ w_\mathrm{PS}^\mathrm{d}(k+1)}.
\label{equ:pi-PS}
\end{equation}  
where the proportionality constant is set by normalisation.

In the following we assume for concreteness that the most likely total occupancy $n^*$ is in the miscibility gap $n_- < (n^*/M) < n_+$, the other cases can be obtained by straightforward extensions.
Then $n^*$ obeys $  w_\mathrm{PS}^\mathrm{b}(n^*)= w_\mathrm{PS}^\mathrm{d}(n^*+1)$, 
so \eqref{equ:pi-PS} yields
\begin{equation}
n^* =  \frac{M[ n_- R(n_+) - n_+ R(n_-)] + w^\mathrm{d}(n_+) - w^\mathrm{d}(n_-)}{R(n_+) - R(n_-)} 
\end{equation}
where $R(n) = w^\mathrm{b}(n) - w^\mathrm{d}(n)$.
Then the fraction appearing in \eqref{equ:pi-PS} can be parameterised for $n_- < (n_{\rm tot}/M) < n_+$ as
\begin{equation}
\label{equ:wPS-AB}
  \frac{ w_\mathrm{PS}^\mathrm{b}(n_{\rm tot})}{ w_\mathrm{PS}^\mathrm{d}(n_{\rm tot}+1)} = \frac{A(n_{\rm tot}-n^*) + B}{n_{\rm tot}-n^* + B}
\end{equation}
where we used \eqref{equ:wPS-sep}, and define
\begin{align}
  A &= \frac{w^\mathrm{b}(n_+) - w^\mathrm{b}(n_-)}{w^\mathrm{d}(n_+) - w^\mathrm{d}(n_-)} \,,
  \\
   B &=  \frac{(M n_+ - n^*) w^\mathrm{b}(n_-) + (n^* -Mn_-) w^\mathrm{b}(n_+)}{w^\mathrm{d}(n_+) - w^\mathrm{d}(n_-)} \,. 
\end{align}

Finally, we take the large-$K$ limit to arrive at a continuous density, as in Sec.~\ref{app:largeK}.  We rescale the total occupancy as $\varphi := n_{\rm tot} / (MK)$.  The microscopic transition rates obey \eqref{equ:wa-K} which means that
$w_{\rm PS}^\alpha(n_{\rm tot}) = (MK) \Lambda_\alpha(n_{\rm tot}/(MK))$ for smooth functions $\Lambda_\alpha$, which ensures that $\varphi$ is typically $O(1)$ at large $K$.
Then following the same arguments as Sec.~\ref{app:FR}, one has from \eqref{equ:pi-PS} that
\begin{equation}
\pi_{\rm PS}(\varphi K) \simeq {\rm e}^{-MK\tilde{F}(\varphi)} 
\end{equation}
with  
\begin{equation}
\label{equ:F-phi}
\tilde{F}(\varphi) = -\int_{\varphi^*}^\varphi \dif\varphi'\, \log\frac{\Lambda_\mathrm{b}(\varphi')}{\Lambda_\mathrm{d}(\varphi')}
\end{equation} 
and $\varphi^* = n^*/(MK)$.  The phase separated regime is $\varphi\in[\rho_-,\rho_+]$ with $\rho_\pm = n_{\pm}/K$, in which case which one finds by (\ref{equ:wa-K},\ref{equ:wPS-AB}) that
$
\Lambda_\mathrm{b}(\varphi)/\Lambda_\mathrm{d}(\varphi) = [A_0(\varphi-\varphi^*) + B_0]/[\varphi-\varphi^* + B_0]
$
with 
\begin{align}
A_0 & = \frac{\lambda^\mathrm{b}(\rho_+) - \lambda^\mathrm{b}(\rho_-)}{\lambda^\mathrm{d}(\rho_+) - \lambda^\mathrm{d}(\rho_-)} \,,
\\
  B_0 & = 
   \frac{( \rho_+ - \varphi^*) \lambda^\mathrm{b}(\rho_-) + (\varphi^* -\rho_-) \lambda^\mathrm{b}(\rho_+)}{\lambda^\mathrm{d}(\rho_+) - \lambda^\mathrm{d}(\rho_-)} \,. 
\end{align}
For the homogeneous cases, (\ref{equ:wa-K},\ref{equ:wPS-hom}) yield $\Lambda_\alpha(\varphi)=  \lambda_\alpha(\varphi)$.
Using these results in \eqref{equ:F-phi} one finally obtains
\begin{align}
\label{eq:F_Poissonian}
  \tilde{F}(\varphi) = \begin{cases}
    \tilde{F}_R(\varphi) - \tilde{F}_R(\rho_-) + \tilde{F}_{\text{PS}}(\rho_-) & \quad \varphi \leq \rho_- \\
    \tilde{F}_{\text{PS}}(\varphi) &  \quad \rho_- < \varphi \leq \rho_+ \\
    \tilde{F}_R(\varphi) - \tilde{F}_R(\rho_+) + \tilde{F}_{\text{PS}}(\rho_+) & \quad \rho_+ < \varphi
 \end{cases}
\end{align}
with  
$\tilde{F}_R(\varphi)$ given by \eqref{eq:F_R} and
\begin{align}
\label{eq:F_PS_Poissonian}
  \tilde{F}_\text{PS}(\varphi) 
  &= -(\varphi - \varphi^*)\log\left( \frac{A_0(\varphi-\varphi^*) + B_0}{\varphi-\varphi^* + B_0}\right) + B_0\log\left( \frac{\varphi-\varphi^*}{B_0} + 1\right)
  -\frac{B_0}{A_0} \log\left( \frac{A_0(\varphi-\varphi^*)}{B_0} +1\right)
  \,.
\end{align}
Compared with the result  \eqref{eq:F_PS} for the model with Gaussianised noise, the function has a similar structure but differs in the details.

Finally, the extinction time in this model behaves analogous to \eqref{eq:djsabdwqbdqdwq}: Repeating the arguments of the main text, one has for large $K$ that
\begin{equation}
\frac{1}{MK} \log \tau  \to   \sup\{ \tilde F(\varphi) :\varphi \in (0,\varphi^*)\}
\end{equation}
where the effective noise strength $(MK)^{-1}$ is proportional to the single-site noise parameter $\epsilon=1/K$ from \eqref{equ:rho-sde} and inversely proportional to the ``system volume'' $M$, similar to \eqref{eq:djsabdwqbdqdwq}.

To summarise this appendix: we constructed a model of particles on a discrete set of lattice sites where the limit of large site occupancy (large $K$) leads to a weak-noise theory for on-site densities [Eq.~\eqref{equ:rho-sde}].  This weak-noise theory relies on a central limit theorem, so the noise is Gaussian.  We analysed  large rare fluctuations of the total density in the discrete model, deriving the free energy $\tilde{F}$, including its behaviour in phase-separated states.  This complements the analysis of the main text, where a similar free energy was computed for the model with Gaussianised noise.  The two models differ  quantitatively because the central limit theorem replaces a Poissonian noise by a Gaussian one, which is accurate for typical fluctuations but fails for rare events.  Nevertheless, the qualitative behaviour of the two models is very similar.  The different results for the two models are summarised in Table~\ref{tab:gauss_vs_poisson}.

\begin{table}[bt]
\def\arraystretch{1.75}
\setlength\tabcolsep{10pt}
  \begin{tabular}{ll|ll}
    Gaussian Noise & & Poissonian Noise \\
    \hline
    $F_A(\rho) = -2\int \dif\rho\, \frac{\lambda_\mathrm{b}(\rho) - \lambda_\mathrm{d}(\rho)}{\lambda_\mathrm{b}(\rho) + \lambda_\mathrm{d}(\rho)}$ 
    & \eqref{eq:F_A} 
    & $F^\mathrm{R}(\rho) = -\int \dif\rho\, \log\frac{\lambda_\mathrm{b}(\rho)}{\lambda_\mathrm{d}(\rho)}$ 
    & \eqref{eq:F_R} 
    \\
    $F(\varphi)$ 
    & \eqref{eq:F} 
    & $\tilde{F}(\varphi)$ 
    & \eqref{eq:F_Poissonian}
    \\
    $F_\text{PS}(\varphi) = -2\int \dif\varphi\, \frac{R_\mathrm{PS}(\varphi)}{M_\mathrm{PS}(\varphi)}$ 
    & \eqref{eq:F_PS} 
    & $\tilde{F}_\text{PS}(\varphi) = -\int \dif\varphi\, \log\frac{\Lambda_\mathrm{PS}^\mathrm{b}(\varphi)}{\Lambda_\mathrm{PS}^\mathrm{d}(\varphi)}$ 
    & \eqref{eq:F_PS_Poissonian}
  \end{tabular}
  \caption{Equivalent quantities differ for Gaussian and Poissonian noise. Results involving Gaussian noise were derived in Sec.~\ref{sec:modelab}, results with Poissonian noise in Appendix.~\ref{sec:discrete_ab}.}
  \label{tab:gauss_vs_poisson}
\end{table}

\section{Mean time to extinction in the toy model}
\label{sec:mte_calc}

In a Markov chain $X_t$ with finite state space $S$, the mean first-passage time to $G \subset S$ starting in $i\in S$ is defined as $\tau(i) = \langle T^G | X_0 = i \rangle$
where the first-passage time is given by $T^G = \inf\{t >0 : X_t \in G \}$. The MTE is the solution to the following system of equations \cite{gardiner2009}
\begin{align}
  0 &= \tau(i) \,,  &\forall i  \in G\,,
  \\
  -1 &= \sum_{j\neq i} r(i\to j) [ \tau(j) - \tau(i) ] \,,  &\forall i  \notin G\,.
\end{align}
For the toy model we define the mean time to extinction (MTE) as the mean first-passage time to 0G, which starting from meso-state $I$ is given by $\tau(I) = \langle \inf\{t >0 : X_t \in \text{0G} \} | X_0 \in I \rangle$. The MTE is explicitly given as the solution of
\begin{align}
    \tau(\text{0G}) &= 0 \,, 
    \\
    -1 &=  r_\text{meso}(\text{2G} \to \text{0G}) (\tau(\text{0G}) - \tau(\text{4G})) - r_\text{meso}(\text{2G} \to \text{4G}) (\tau(\text{2G}) - \tau(\text{4G})) \,, 
    \\
    -1 &= r_\text{meso}(\text{4G} \to \text{2G}) (\tau(\text{2G}) - \tau(\text{4G}))\,,
\end{align}
which is
\begin{align}
  \tau(\text{2G}) &= \frac{1}{r_\text{meso}(\text{2G} \to \text{0G}) } \frac{r_\text{meso}(\text{2G} \to \text{4G})}{r_\text{meso}(\text{4G} \to \text{2G})} + \frac{1}{r_\text{meso}(\text{2G} \to \text{0G})} \,,
  \\
  \tau(\text{4G}) &= 
  \frac{1}{r_\text{meso}(\text{2G} \to \text{0G}) } \frac{r_\text{meso}(\text{2G} \to \text{4G})}{r_\text{meso}(\text{4G} \to \text{2G})} 
  + \frac{1}{r_\text{meso}(\text{2G} \to \text{0G})} 
  + \frac{1}{r_\text{meso}(\text{4G} \to \text{2G})} \,.
  \label{eq:tau34raw}
\end{align}
Using detailed balance (see Eqs.  (\ref{equ:toy-meso-db},\ref{eq:fnjnojwqhiwqeewq}) in the main text) we find that
\begin{equation}
\label{eq:danpcngpwk}
  \lim_{\beta\to \infty} \beta^{-1} \log \frac{r_\text{meso}(J \to I)}{r_\text{meso}(I \to J)} = f(J) - f(I) \, .
\end{equation}
For the toy model one may additionally verify from \eqref{eq:rmeso} that
\begin{equation}
  \lim_{\beta\to \infty} \beta^{-1} \log k/r_\text{meso}(I \to J) = \max\{0, f(J) - f(I)\} \,.
\end{equation}
By the same argument as in \eqref{equ:r-sup} of the main text\footnote{
For a set of functions $A_1(\beta),A_2(\beta), \ldots A_N(\beta) > 0$ denote $a_i = \lim_{\beta \to \infty} \beta^{-1} \log A_i(\beta)$ and suppose that all these limits exist. It then holds that 
$\lim_{\beta \to \infty} \beta^{-1} \log \sum_{i=1}^N A_i (\beta) = \max_i\left\{ a_i \right\}$. 
}
we see that $\beta^{-1} \log [k \tau(\text{4G})]$ is dominated at low temperatures by the largest term on the right hand side of \eqref{eq:tau34raw}.  Recalling that $f(4\GG)>0$, the relevant limits are
\begin{align}
  \lim_{\beta \to \infty} \beta^{-1} \log \left( k  \frac{r_\text{meso}(\text{2G} \to \text{4G})}{r_\text{meso}(\text{2G} \to \text{0G}) r_\text{meso}(\text{4G} \to \text{2G})} ) \right) &= \max\{0, f(\text{0G}) - f(\text{2G})\} + f(\text{2G}) \,,
  \\
  \lim_{\beta \to \infty} \beta^{-1} \log k /r_\text{meso}(\text{2G} \to \text{0G}) &= \max\{0, f(\text{0G}) - f(\text{2G})\} \,,
  \\
  \lim_{\beta \to \infty} \beta^{-1} \log k /r_\text{meso}(\text{4G} \to \text{2G}) &= \max\{0, f(\text{2G})\} \,.
\end{align}
Taking the max of these three terms and restricting as in the main text that 4G is a local minimum of $f$ [so $f(\text{2G})>f(\text{4G}) =0$]  we obtain
\begin{equation}
  \lim_{\beta\to \infty} \beta^{-1} \log [k \tau(\text{4G})] = \max\{f(\text{0G}) , f(\text{2G}) \} \,.
\end{equation}
Hence identifying $\tau_\text{ext} = \tau(\text{4G})$ we recover Eq.~\eqref{equ:def-f*} of main text.